\documentclass[namedreferences]{solarphysics}
\usepackage[hyperref,optionalrh,solaromanenum]{spr-sola-addons} % For Solar Physics
\usepackage{epsfig}                     % For eps figures, old commands
\usepackage{graphicx}                    % For eps figures, newer & more powerfull
\usepackage{amssymb}                    % useful mathematical symbols
\usepackage{color}                       % For color text: \color command
\usepackage{breakurl}                         % For breaking URLs easily trough lines
\usepackage{subfig}
\newcommand{\LL}{\left(}
\newcommand{\RR}{\right)}
\newcommand{\LS}{\left[}
\newcommand{\RS}{\right]}

\newcommand{\D}[2]{\frac{d#1}{d#2}}
\renewcommand{\P}[2]{\frac{\partial #1}{\partial #2}}

\newcommand{\bb}[1]{\textsl{\textbf{#1}}}

\newcommand{\pr}{\parallel}
\newcommand{\g}{\gamma}
\newcommand{\ex}{\textrm{e}}
\renewcommand{\a}{\alpha}
\renewcommand{\b}{\beta}
\renewcommand{\deg}{^{\circ}}

\DeclareRobustCommand*{\unit}[1]{\def~{\,}\ensuremath{\mathrm{\,#1}}}
\chardef\us=`\_

\begin{document}

\begin{article}

\begin{opening}

\title{Particle Acceleration in Collapsing Magnetic Traps with a Braking Plasma Jet}

%%%%%%%%%%%%%%%%%%%%%%%%%%%%%%%%%%%%%%%%%%%%%%%%%%%
% Authors Names

\author[addressref={aff1},corref,email={ab325@st-andrews.ac.uk}]{\inits{A.}\fnm{Alexei}~\lnm{Borissov}}
\author[addressref={aff1},email={tn3@st-andrews.ac.uk}]{\inits{T.}\fnm{Thomas}~\lnm{Neukirch}}
\author[addressref={aff1},email={jwt9@st-andrews.ac.uk}]{\inits{J.}\fnm{James}~\lnm{Threlfall}}
\address[id=aff1]{School of Mathematics and Statistics, University of St Andrews, St Andrews KY16 9SS, UK}

%%%%%%%%%%%%%%%%%%%%%%%%%%%%%%%%%%%%%%%%%%%%%%%%%%%
%% Runningheads
%
%\runningauthor{}
%\runningtitle{}

%%%%%%%%%%%%%%%%%%%%%%%%%%%%%%%%%%%%%%%%%%%%%%%%%%%
%% Affilations
%% id shold be the same with \author addressref value.
%\address[id={}]{}

%%%%%%%%%%%%%%%%%%%%%%%%%%%%%%%%%%%%%%%%%%%%%%%%%%%
%%% Abstract
\begin{abstract}
Collapsing magnetic traps (CMTs) are one proposed mechanism for generating non-thermal particle populations in solar flares. CMTs occur if an initially stretched magnetic field structure relaxes rapidly into a lower-energy configuration, which is believed to happen as a by-product of magnetic reconnection. A similar mechanism for energising particles has also been found to operate in the Earth's magnetotail. One particular feature proposed to be of importance for particle acceleration in the magnetotail is that of a braking plasma jet,  \textit{i.e.} a localised region of strong flow encountering stronger magnetic field which causes the jet to slow down and stop. Such a feature has not been included in previously proposed analytical models of CMTs for solar flares. In this work we incorporate a braking plasma jet into a well studied CMT model for the first time. We present results of test particle calculations in this new CMT model. We observe and characterise new types of particle behaviour caused by the magnetic structure of the jet braking region, which allows electrons to be trapped both in the braking jet region and the loop legs. We compare and contrast the behaviour of particle orbits for various parameter regimes of the underlying trap by examining particle trajectories, energy gains and the frequency with which different types of particle orbit are found for each parameter regime.
\end{abstract}

%%%%%%%%%%%%%%%%%%%%%%%%%%%%%%%%%%%%%%%%%%%%%%%%%%%
%% Keywords
%
%\keywords{}

\end{opening}
%-------------------------------------------------

%%%%%%%%%%%%%%%%%%%%%%%%%%%%%%%%%%%%%%%%%%%%%%%%%%%%%%%%%%%%%%%%%%%%%%%%%%%%%%%%%%%%%%%%%%%%%%%%%%%%%%%%%%%%%%%%%%%%%%%%

\section{Introduction}

Magnetic reconnection is thought to play an important role in releasing energy stored in magnetic fields in multiple environments, including solar flares \citep[\textit{e.g.}][]{zharkova-et-al2011} and substorms in Earth's magnetotail \citep[\textit{e.g.}][]{birn-et-al2012}. In situ observations \citep[\textit{e.g.}][]{imada-et-al2007,fu-et-al2013} and numerical simulations \citep[\textit{e.g.}][]{sitnov-swisdak2011} also suggest that jets with abrupt fronts characterized by large increases in magnetic field strength may be caused by magnetic reconnection. Interaction of the jet with stronger fields causes the jet to slow down and stop \citep[\textit{e.g.}][]{khotyaintsev-et-al2011}. This effect is called jet braking and has been shown to accelerate particles \citep{artemyev2014}.

A major open question in the physics of solar flares lies in describing the mechanisms responsible for accelerating particles \citep[for an overview of this topic see reviews by][]{miller-et-al1997,zharkova-et-al2011,cargill-et-al2012}. A variety of mechanisms have been proposed, including acceleration by parallel electric field in the reconnection region \citep[\textit{e.g.}][]{litvinenko1996,wood-neukirch2004,gordovskyy-et-al2010a,gordovskyy-et-al2010b}, acceleration at shocks \citep[\textit{e.g.}][]{cargill1991,tsuneta-naito1998,miteva-mann2007,mann-warmuth2009,chen-et-al2015}, and stochastic acceleration \citep[\textit{e.g.}][]{miller-et-al1997}. 

It has also been shown that particle acceleration and trapping may be obtained in what is known as a collapsing magnetic trap (CMT) \citep[\textit{e.g.}][]{somov-kosugi1997,bogachev-somov2001,bogachev-somov2005,bogachev-somov2007,bogachev-somov2009,somov-bogachev2003,karlicky-kosugi2004,karlicky-barta2006,grady-neukirch2009,minoshima-et-al2010,minoshima-et-al2011,grady-et-al2012,eradat_oskoui-et-al2014}. Generally speaking, this involves a stretched magnetic field prior to the onset of reconnection relaxing to a lower energy configuration. Particles gain energy in this scenario due to betatron and Fermi acceleration.

We propose a model to combine a braking jet with a CMT and investigate trapping and acceleration mechanisms generated by such a trap configuration in the context of solar flares. In this model a plasma jet produced at the reconnection region in the standard model of solar flares \citep[see][]{shibata-magara2011} propagates sunward and forms a braking jet as it enters regions of higher magnetic field closer to the solar surface. This is accompanied by a relaxation of the field lines analogous to a CMT. In contrast to other models of CMTs, our proposed model incorporates a front associated with the braking jet at which there is a pileup of magnetic flux. 

Due to the propagation of the front and the high magnetic field strength associated with it, a patch of relatively strong perpendicular electric field (in comparison with a normal CMT) is produced in the region of the front. This is consistent with magnetohydrodynamic (MHD) simulations by, for example, \citet{birn-hesse1996} in the context of the Earth's magnetotail, as well as for solar flares by \citet{karlicky-barta2006}. In addition to the front propagation, we add an indentation into the front caused by the interaction of the braking plasma jet with the low-lying magnetic field loops. This structure is also seen in the MHD simulations in \cite{birn-hesse1996} and \cite{karlicky-barta2006}. Finally, \citet{artemyev2014} suggests that as the front propagates it may get steeper (in the sense that there is more pileup of magnetic field lines). This steepening of the front is incorporated into our model at the beginning of the simulation when the jet is propagating through regions of lower magnetic field strength. As the jet encounters regions of strong field and slows down sufficiently we reduce the pileup of magnetic field lines. 
 
The aim of this work is to study charged particle dynamics in such a model. To that end our paper is organised as follows: in Section~\ref{transformation} we incorporate a braking jet into the CMT model of \cite{giuliani-et-al2005}. To do this, we modify the analytical expressions for the electromagnetic fields describing a CMT to include a front and a braking jet. We initialize test particle orbits with a wide range of initial conditions in these analytical fields and solve the relativistic guiding centre equations for the particle orbits numerically. In this paper we only examine the orbits of electrons. In Section~\ref{sample-trajectories} we present several cases of typical particle orbits seen in this model. Section~\ref{particle-motion} discusses the key factors underpinning the different behaviour observed.  Finally in Section~\ref{frequencies} we discuss the relative abundance of different types of particle orbit behaviour for various model parameters, and any resulting effects on test particle energization. 

%%%%%%%%%%%%%%%%%%%%%%%%%%%%%%%%%%%%%%%%%%%%%%%%%%%%%%%%%%%%%%%%%%%%%%%%%%%%%%%%%%%%%%%%%%%%%%%%%%%%%%%%%%%%%%%%%%%%%%%%

\section{Analytical Model of Electromagnetic Fields}\label{transformation}

To define the electromagnetic fields analytically we follow the transformation method of \citet{giuliani-et-al2005}. This approach assumes that the evolution of the CMT occurs in a region of the corona where ideal MHD holds. Ideal MHD implies that the magnetic field is frozen to the plasma flow, which means that the fields may be calculated from a specified flow. Integration of the plasma flow velocity leads to an expression for the position of a fluid element at the final time (at the end of the simulation) as being a function of the position of the fluid element at an earlier time, that is $\bb x_\infty = \bb G(\bb x,t)$. We choose our model to be two-dimensional, so we can write 
\begin{equation}\label{B-def}
\bb B = \nabla A (x,y,t) \times \bb e_z + B_z \bb e_z .
\end{equation}
We specify the flux function, $A$, (and hence the electromagnetic fields) at $(\bb x, t)$ by choosing a final flux function, $A_0(\bb x)$ and a coordinate transformation $\bb x_\infty(\bb x, t)$ so that $A(\bb x,t) = A_0\LS\bb x_\infty(\bb x,t)\RS$. In the present investigation we set $B_z = 0$.

Following \citet{giuliani-et-al2005}, we choose a bipolar configuration for the final flux function of the form
\begin{equation}
A_0(\bb x) = c_1\LS \arctan\LL\frac{y + 1}{x + 0.5}\RR + \arctan \LL\frac{y + 1}{x - 0.5} \RR \RS,
\end{equation}
where $x$ is the spatial coordinate parallel to the solar surface, $y$ is the coordinate normal to the solar surface, and $c_1$ determines the strength of the bipole. As in \citet{giuliani-et-al2005}, we choose the normalizing length scale to be $\hat L = 10^7\unit{m}$.
We choose a coordinate transformation given by:

\begin{equation}\label{yinf}
	y_\infty = \LS s\log \LL 1 + \frac{y}{s} \RR  \LL 1 - F(y)G(x) \RR + y F(y) G(x)  + y\frac{1 + \tanh \phi}{2}\RS f(t) + y g(t),
\end{equation}
\begin{equation}\label{xinf}
	x_\infty = x,
\end{equation}
where 
\[f(t) = \frac{1 - \tanh(t - t_0)}{2},\quad g(t) = 1 - f(t),\]
\[F(y) = \frac{1 - \tanh(y - y')}{2}, \quad G(x) = \frac{\tanh(x + x') - \tanh(x - x')}{2},\]
\begin{eqnarray} \label{phi}
\phi &=& \a\LL 1 + \LL \chi \sin\LL \frac{\pi y}{y_0} \RR - 1 \RR\frac{1 - \tanh(\zeta (t - t'))}{2} \RR \\
&& \cdot\LL \LL y + 1 \RR \LL s_{o} x^{2} + 1 \RR  \RR ^\b \LL y - v_\phi \sigma \tanh \LL \frac{t}{\sigma} \RR - y_0 + J \RR + T, \nonumber
\end{eqnarray}
and 
\begin{equation} \label{jet}
J = d \ex^{-x^2 y /w},
\end{equation}
\begin{equation}\label{trap}
T = k\LL 1 + \LL \chi \sin\LL \frac{\pi y}{y_0} \RR - 1 \RR\frac{1 - \tanh(\zeta (t - t'))}{2} \RR\tan \LL \frac \pi 2 \frac{x^2}{w_2} \RR \tanh (y).
\end{equation}
The Appendix contains a more detailed discussion of how the transformation in Equations~(\ref{yinf}) -- (\ref{trap}) is obtained.

The properties of the CMT are determined by the parameters in the transformation in Equations~(\ref{yinf}) -- (\ref{trap}). The parameters $d,w,\b,w_2,k,x',y',s$ and $s_o$ are chosen so that the shape of the field lines resembles that found in MHD simulations, for example by \cite{karlicky-barta2006}. The steepness of the front is controlled by $\alpha, \chi, \zeta$ and $t'$. These parameters are chosen so that the gradient of the magnetic field at the jet braking region is initially increasing (when magnetic flux is piling up), and later decreasing (when the jet is being slowed by the stronger magnetic field in the lower loops). Of the remaining parameters, the initial speed of the front is determined by $v_\phi$, the initial position of the front by $y_0$, $\sigma$ controls the distance to which the front propagates, and the time at which the front dissipates is given by $t_0$.

The value of $\a$ is set by balancing the deceleration of the jet with deceleration forces due to both curvature of the field lines at the centre of the braking jet region and magnetic pressure. In the frame of reference of the jet the acceleration is $\partial v/ \partial t$. By assuming a normalising timescale of $\hat T = 10\unit{s}$, the acceleration in the frame of reference of the jet is approximately $\hat L/\hat T ^2 = 10^5\unit{m}~\unit{s^{-2}}$. This can also be related to the forces acting on the plasma in the jet as follows 

\begin{equation}\label{decel}
\P{v}{t} \approx \frac{1}{\rho}\LL \frac{B^{2}}{\mu_{0} R_{c}} - \nabla \frac{B^{2}}{2\mu_{0}}\RR.
\end{equation}
The two terms on the right hand side of Equation~(\ref{decel}) are the magnetic tension and pressure terms respectively. Assuming a mass density of $\rho = 10^{-10}~\unit{kg}~\unit{m^{-3}}$, a magnetic field of order $\hat B = 10^{-2}~\unit{T}$ in the CMT, a radius of field line curvature $R_{c} \approx 10^{7}\unit{m}$ in the jet braking region, and approximating the magnetic permeability as $\mu_{0} \approx 10^{-6}~\unit{H}~\unit{m^{-1}}$, requires a magnetic field gradient $\nabla \LL B^{2} \RR \approx 10^{-11}~\unit{T^2}~\unit{m^{-1}}$. 

\begin{figure}[h!]
	\includegraphics[trim = 0cm 3cm 0cm 0cm,clip = true,width = \textwidth]{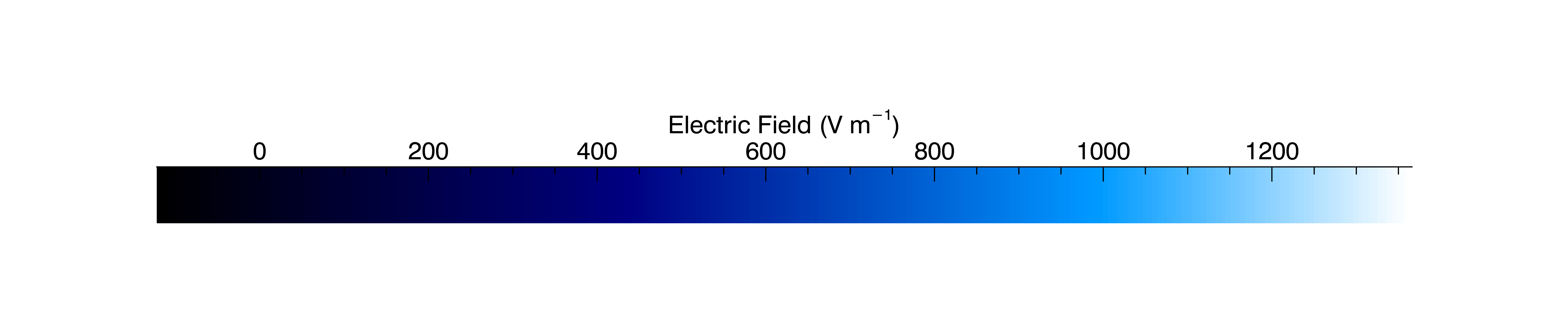}
	\\
	\subfloat[][$t = 0\unit{s}$]{\includegraphics[width = 0.32\textwidth]{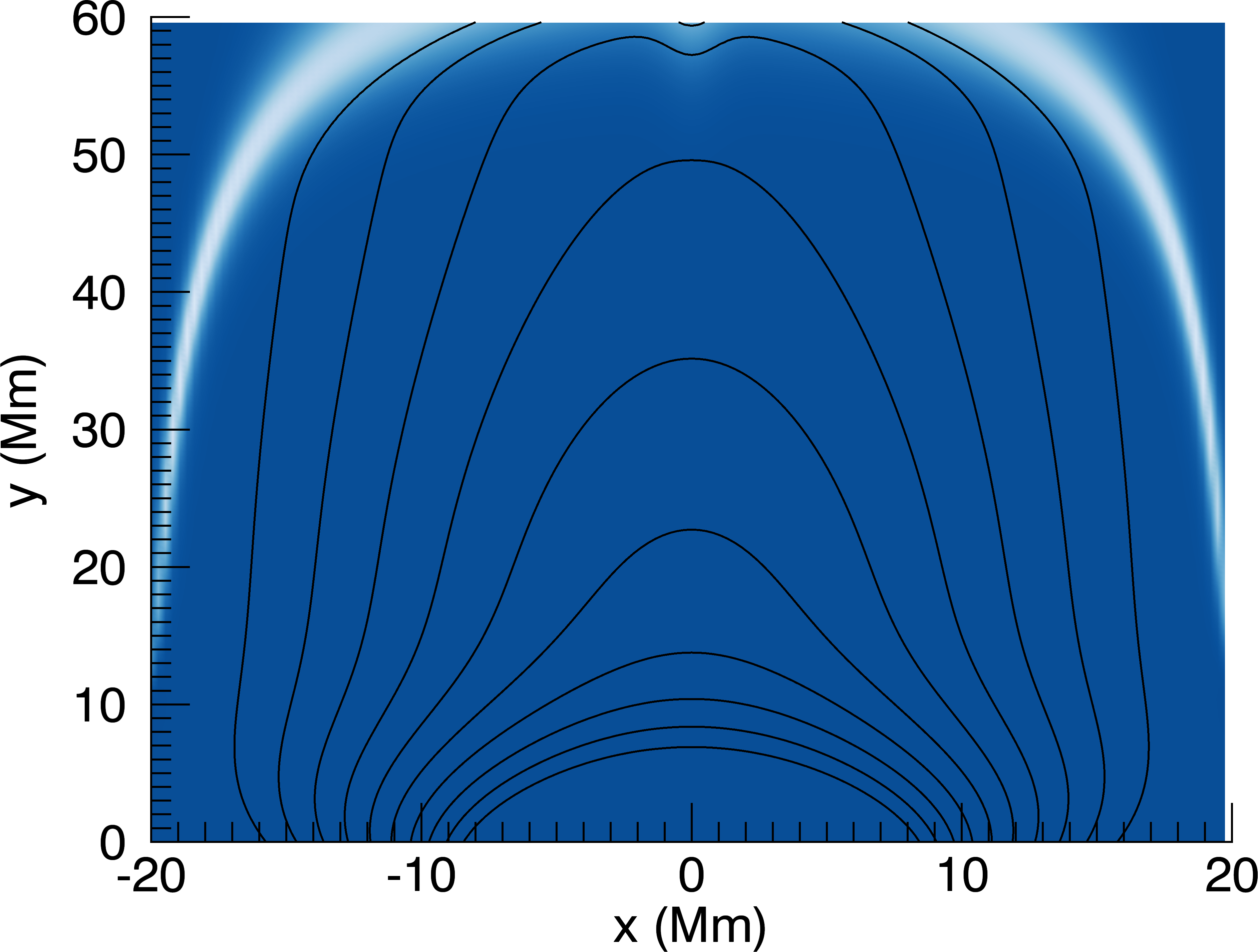}}
	\subfloat[][$t = 5\unit{s}$]{\includegraphics[width = 0.32\textwidth]{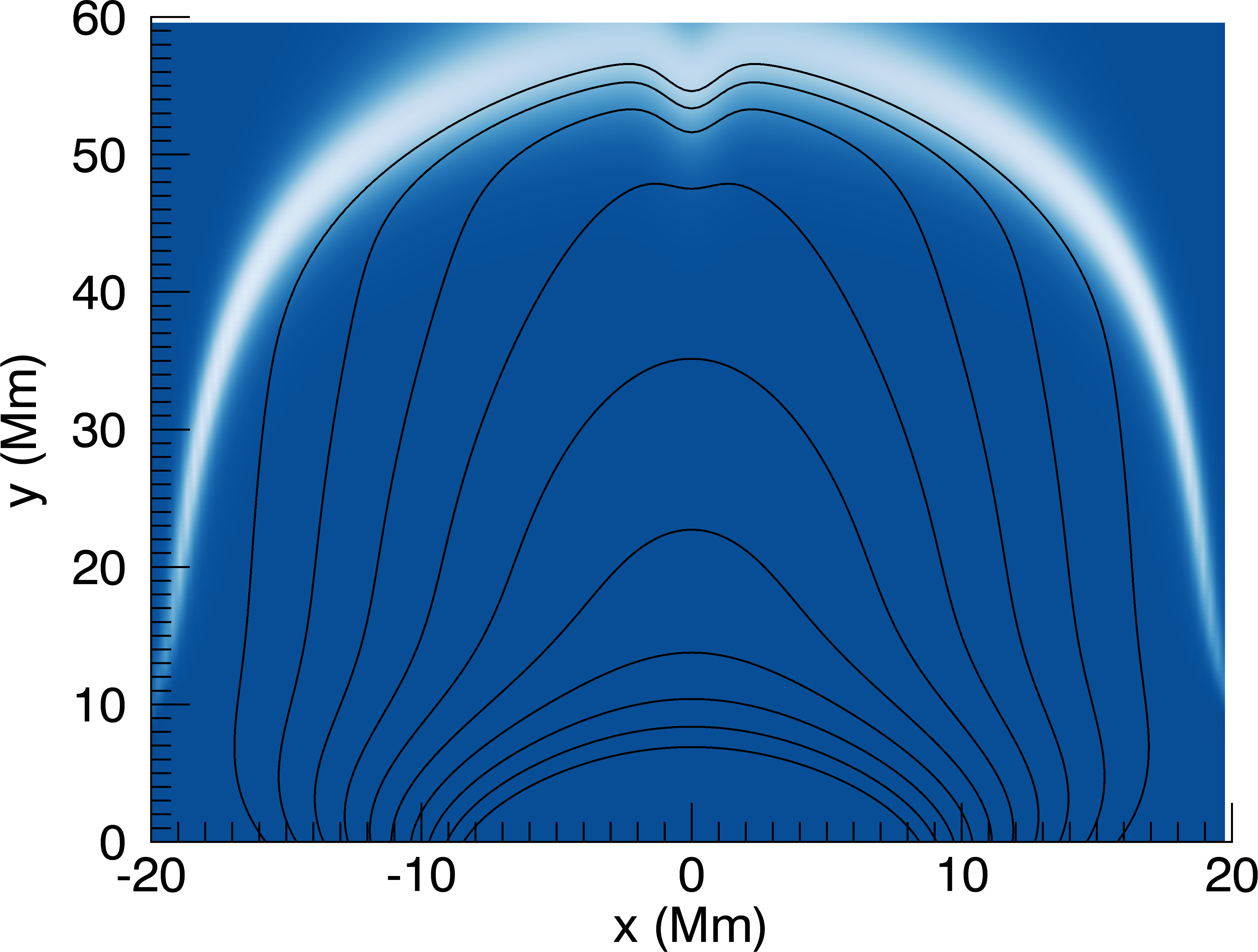}}
	\subfloat[][$t = 10\unit{s}$]{\includegraphics[width = 0.32\textwidth]{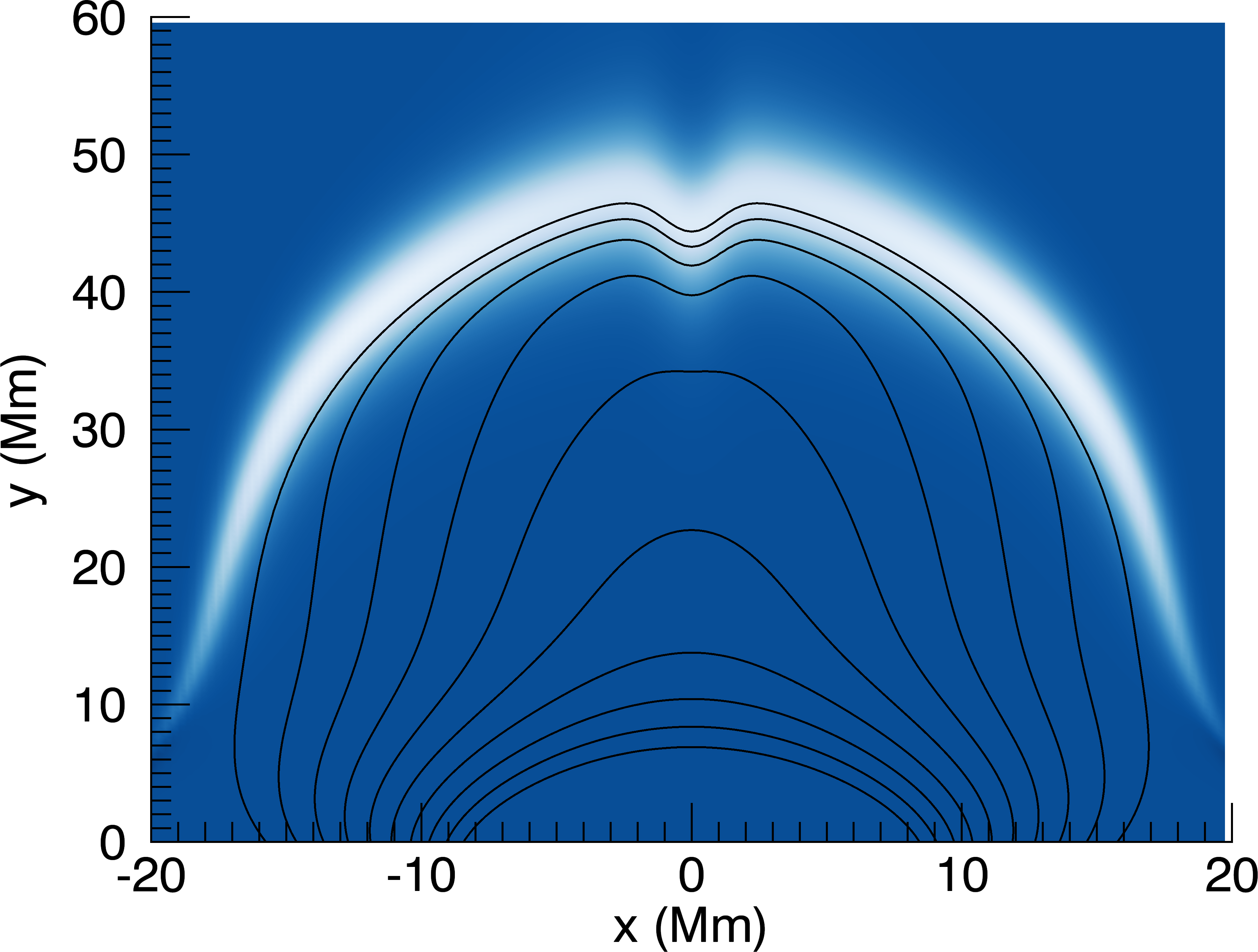}}
	\\
	\subfloat[][$t = 15\unit{s}$]{\includegraphics[width = 0.32\textwidth]{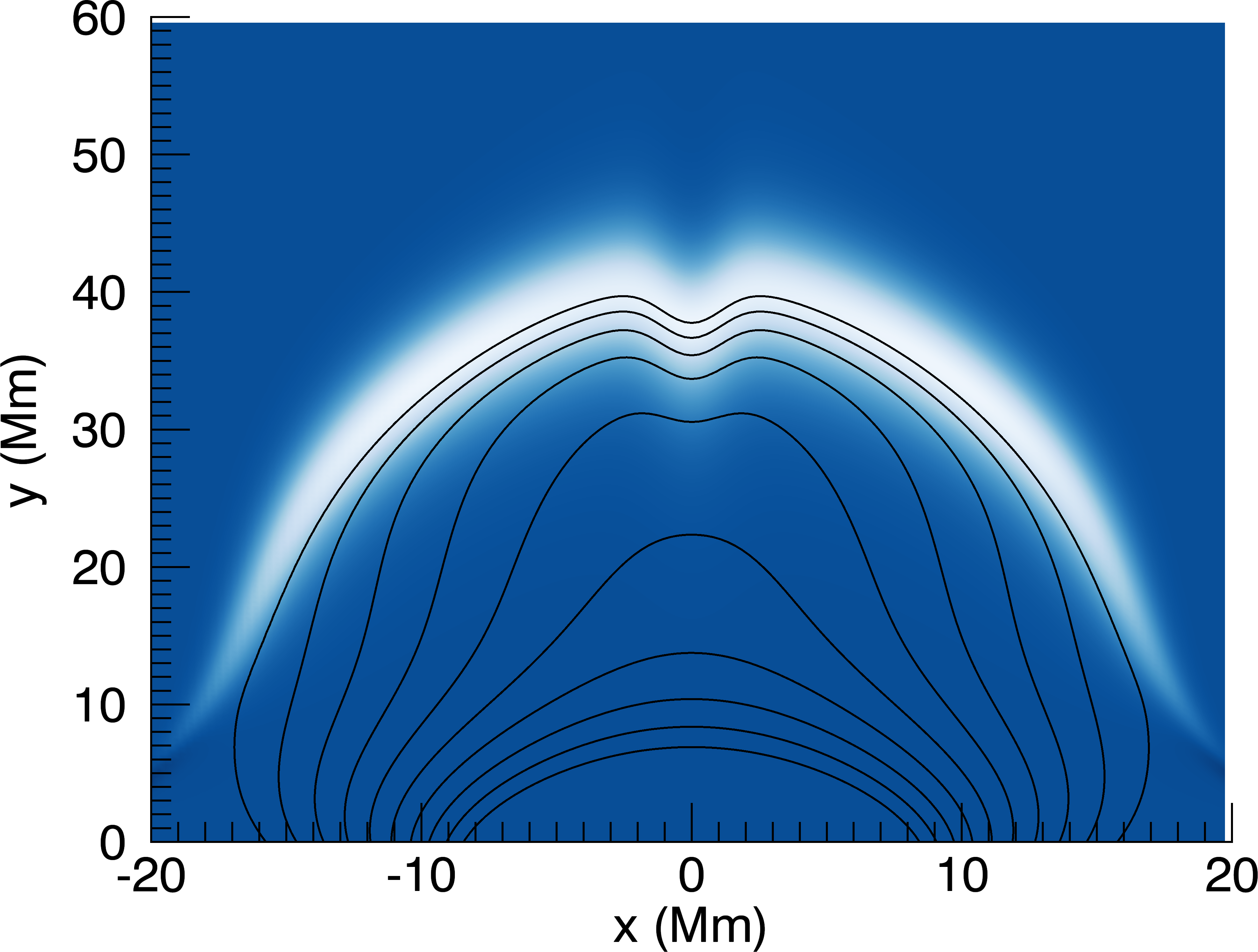}}
	\subfloat[][$t = 20\unit{s}$]{\includegraphics[width = 0.32\textwidth]{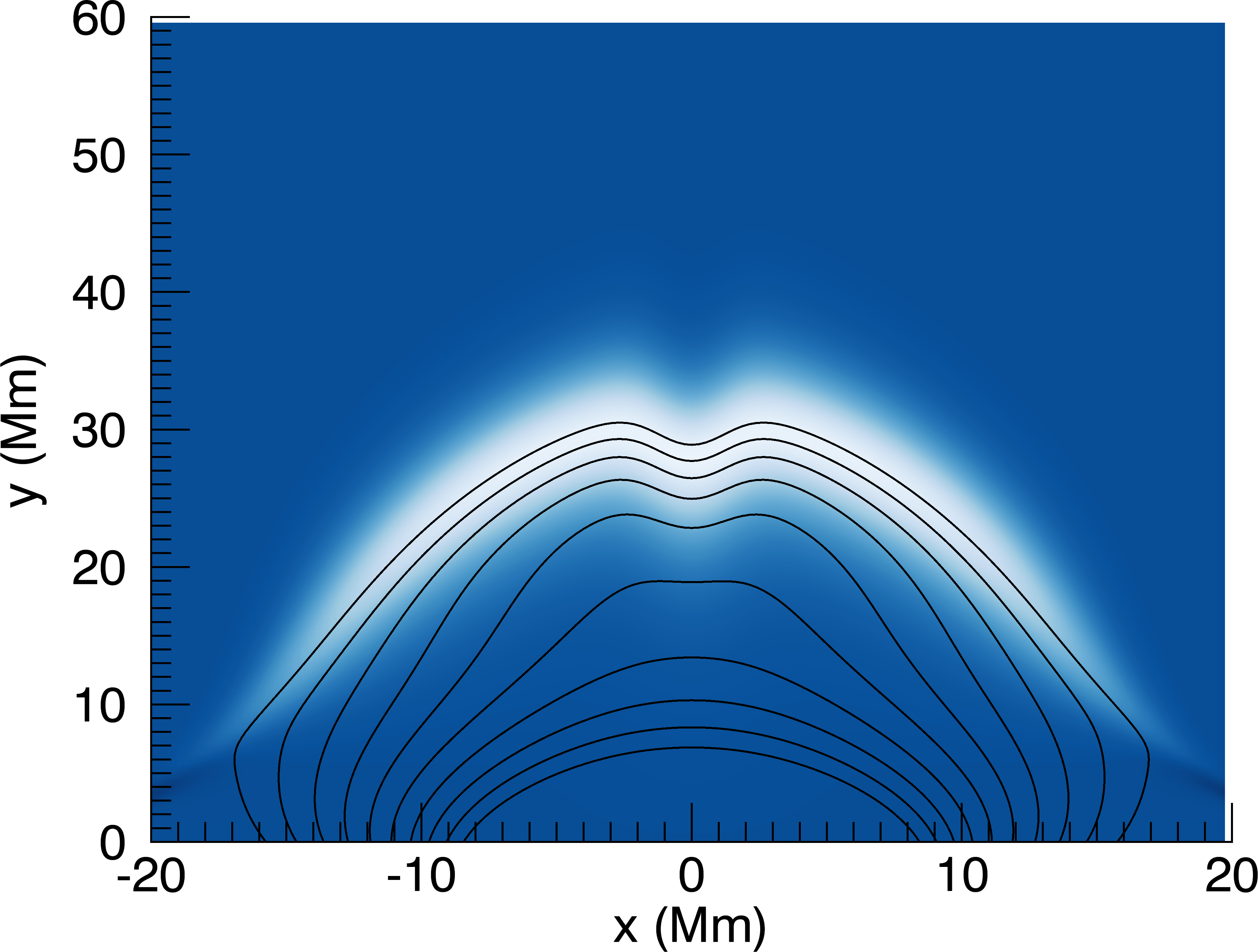}}
	\subfloat[][$t = 25\unit{s}$]{\includegraphics[width = 0.32\textwidth]{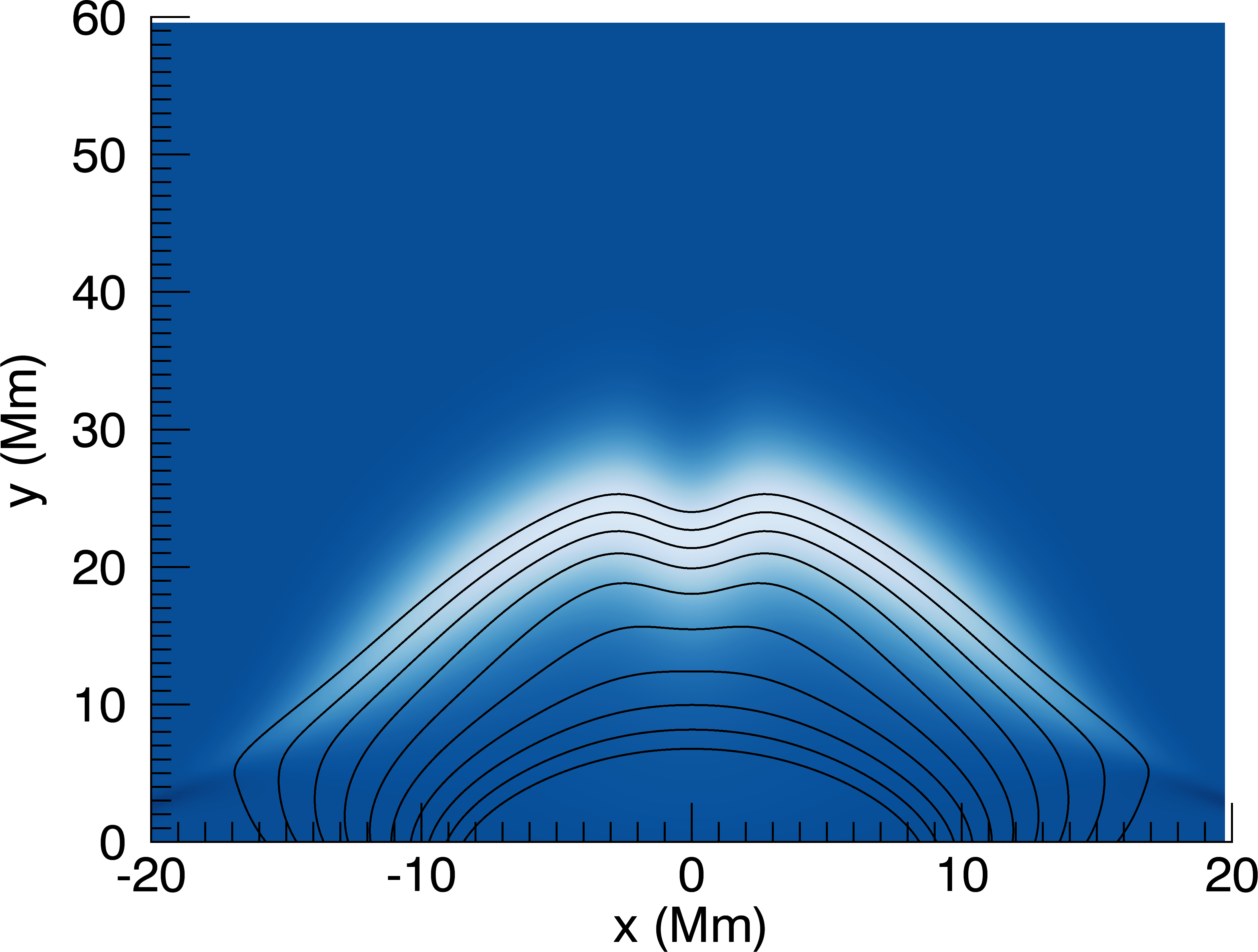}}
	\\
	\subfloat[][$t = 30\unit{s}$]{\includegraphics[width = 0.32\textwidth]{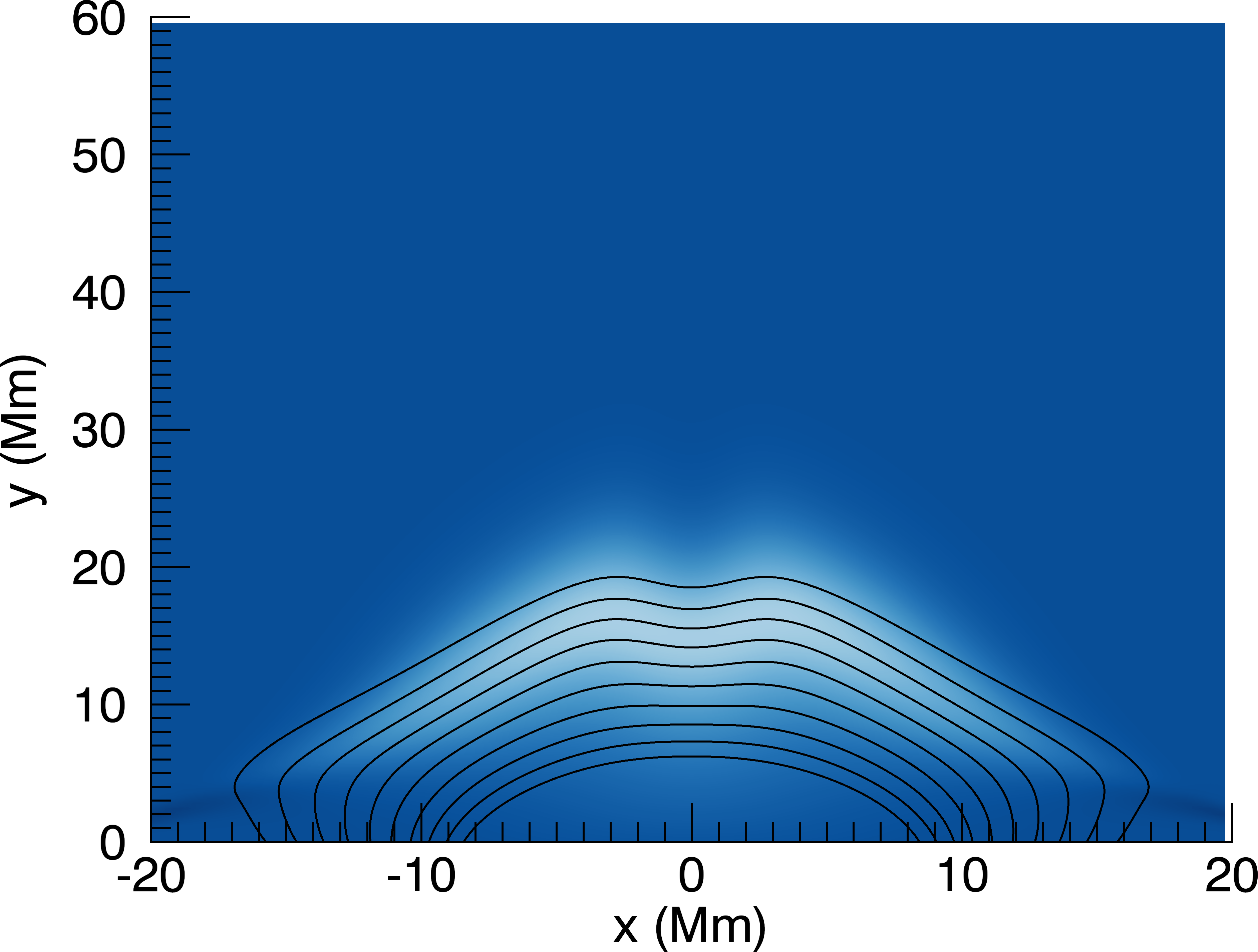}}
	\subfloat[][$t = 35\unit{s}$]{\includegraphics[width = 0.32\textwidth]{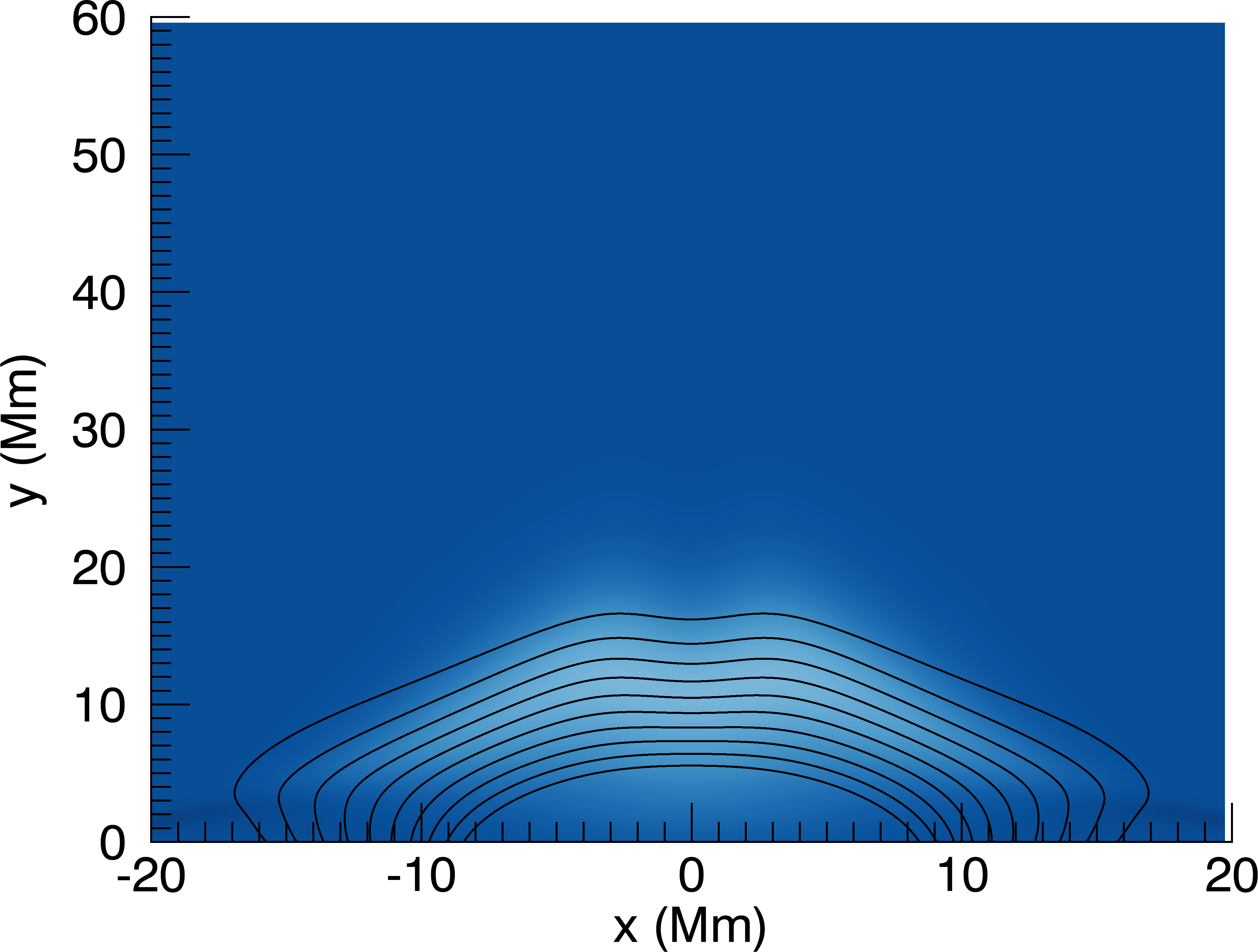}}
	\subfloat[][$t = 40\unit{s}$]{\includegraphics[width = 0.32\textwidth]{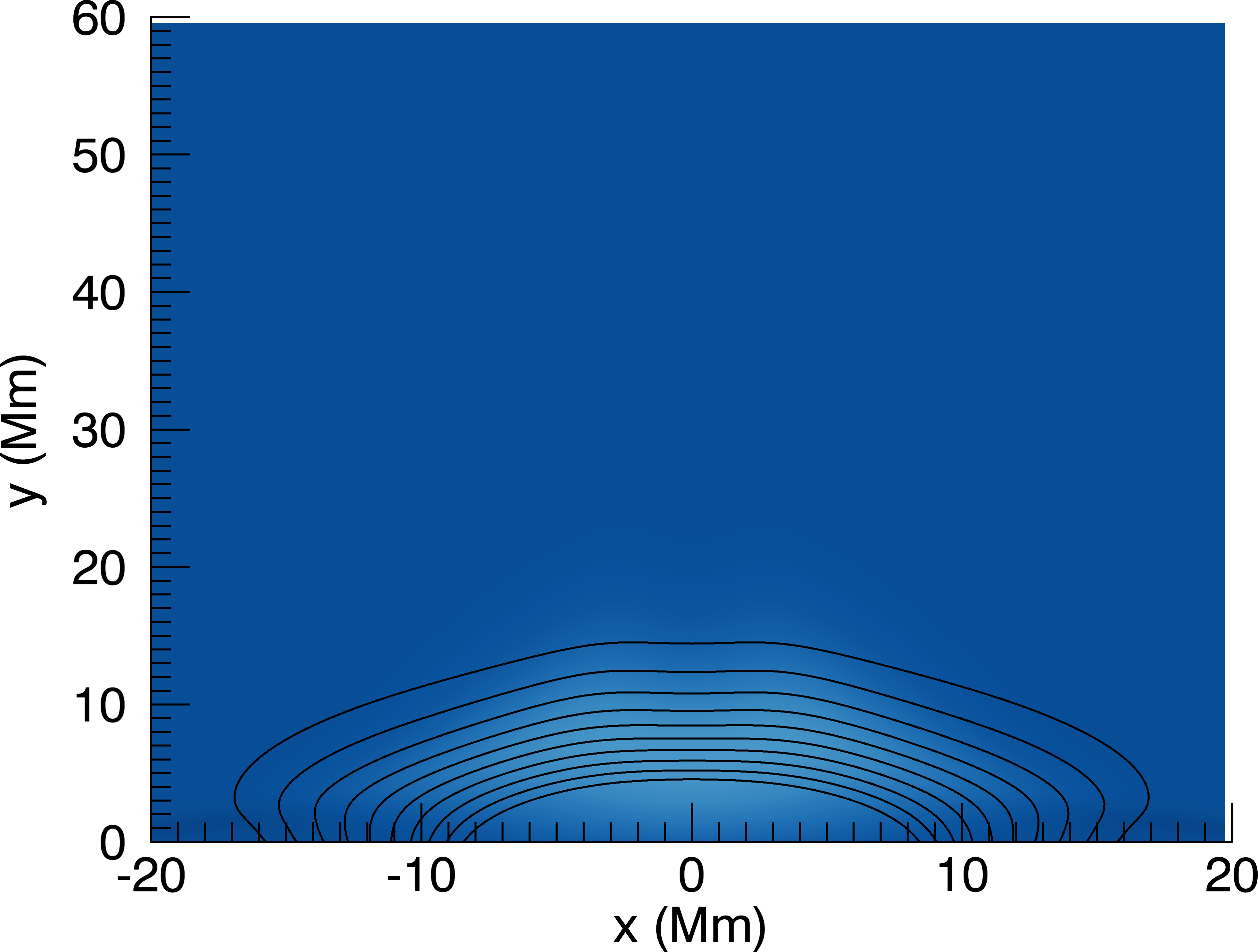}}
	\caption{Panels a-i outline the temporal evolution of a collapsing magnetic trap which incorporates jet braking and flux pileup. The evolution of the electric field is seen in colour and overlaid with magnetic field lines (black) to illustrate the evolution of the magnetic field.}
	\label{em-fields}
\end{figure}

One set of non-dimensional parameters reproducing the properties of the CMT model we require is given in Table \ref{parameters}. In this paper we refer to this set of parameters as the basic parameters. The magnetic field lines and out-of-plane electric field given by our CMT model using the basic parameters at various times are shown in Figure~\ref{em-fields}. The desired features, in particular, the pileup of magnetic flux, the indentation due to the braking jet deforming field lines, and the slowing and dissipation of the jet front, are clearly evident in Figure~\ref{em-fields}. Extraneous patches of strong electric field for large values of horizontal distance from the centre of the trap ($|x| \gtrsim 15\unit{Mm}$) and early times ($t < 1\unit{s}$) are undesirable, however, they do not affect particle orbits because test particles are initialized on field lines that do not cross this region.

\begin{table}
\begin{tabular}{*{14}{c}}
\hline
 \multicolumn{4}{c}{Shape} & \multicolumn{6}{c}{Front structure} & \multicolumn{4}{c}{Front propagation} \\
\hline
	$d$ & 0.3 & $k$ & 7 &&& $\a$ & 1 &&& $\sigma$ & 4 & $x'$ & 100 \\ $w$ & 0.1 & $s$ & 0.7 &&& $\chi$ & 1 &&& $v_\phi$ & -2 & $y'$ & 1 \\ $w_2$ & 2.3 & $s_0$ & 0.5 &&& $\zeta$ & 0.3 &&& $y_0$ & 7 & $t'$ & 5 \\ $\beta$ & 0.5 &  &  & &&& & & &&& $t_0$ & 100 \\
\hline
\end{tabular} 
\caption{Basic non-dimensional parameters used in the CMT model (Equations~(\ref{yinf})-(\ref{trap})) which control the shape of the trap, the steepness of the front, and the propagation of the front. The Appendix details how these parameters affect the CMT model.}
\label{parameters}
\end{table}

%%%%%%%%%%%%%%%%%%%%%%%%%%%%%%%%%%%%%%%%%%%%%%%%%%%%%%%%%%%%%%%%%%%%%%%%%%%%%%%%%%%%%%%%%%%%%%%%%%%%%%%%%%%%%%%%%%%%%%%%

\section{Sample Test Particle Orbits}\label{sample-trajectories}
The particle trajectories are solved for by integrating the relativistic guiding centre equations as given in \citet[][p.32]{northrop1963} \citep[for similar numerical implementations of these equations see][]{gordovskyy-et-al2010b,eradat_oskoui-neukirch2014,threlfall-et-al2015}. The equations are:

\begin{equation}\label{NRGC1}
\dot{\bb R}_{\perp} = \frac{\bb b}{B} \times \LS -\bb E + \frac{\mu}{\g e} \nabla B + \frac{mU}{e} \D{\bb b}{t} + \frac{m\g}{e}\D{\bb u_{E}}{t} + \frac{U}{\g}E_{\pr}\bb u_{E} + \frac{\mu}{\g e} \bb u_{E} \P{B}{t} \RS,
\end{equation}

and

\begin{equation}\label{NRGC2}
m\D{U}{t} = m\g \bb u_{E} \cdot \D{\bb b}{t} + eE_{\pr} - \frac{\mu}{\g}\P{B}{s},
\end{equation}

where $\gamma$ is given by:

\begin{equation}\label{NRGC3}
\g = \sqrt{1 + \frac{U^2 + u_E^2}{c^2} + \frac{2\mu B}{mc^2}}.
\end{equation}
Here $\bb u_E = \bb E \times \bb B/B^2$ is the E cross B drift velocity of the guiding centre, and $U = \g v_\pr$, where $v_\pr = \bb v \cdot \bb b$ and $\bb b$ is the unit vector in the direction of the magnetic field. $\bb R$ refers to the position of the guiding centre, so $\dot{\bb R}_\perp$ is the guiding centre drift velocity perpendicular to the magnetic field. Lastly, $\mu = p_\perp^2/2mB$ is the relativistic magnetic moment, where $p_\perp = \g m v_\perp$ is the perpendicular momentum of the particle, $m$ is the electron mass, $e$ is the electron charge and $c$ is the speed of light. The guiding centre equations given in \cite{northrop1963} contain factors of $\LL 1 - E^2/B^2 \RR$ which are set to unity in Equations~(\ref{NRGC1})--(\ref{NRGC2}) due to our assumption that the plasma flow is non-relativistic.

Equations~(\ref{NRGC1})--(\ref{NRGC3}) are normalized (subject to the values of $\hat L, \hat T $ and $\hat B$ given in Section~\ref{transformation}) and solved numerically using an adaptive timestep Runge-Kutta scheme \citep[specifically the Cash-Karp method, see][for details]{cash-karp1990}. This method adjusts the timestep of the code so that the difference between 4th and 5th order Runge-Kutta solutions is below a chosen tolerance. Since Equation~(\ref{NRGC3}) defines $\g$, it does not need to be integrated, however it is used to update the value of $\g$ at the beginning of each timestep. Test particle orbits are initialized in the CMT with a specified position, kinetic energy and pitch angle, $\theta \equiv \arctan \LL v_\perp/v_\pr\RR$.  In this paper we refer to pitch angles near $\theta = 90\deg$ as high pitch angles, and pitch angles near $\theta = 0\deg$ or $\theta = 180 \deg$ as low pitch angles. We illustrate the most common particle trajectories in Sections \ref{type-1}--\ref{type-3}. Section~\ref{type-1} describes particle orbits which are dominated by trapping in the jet braking region near the centre of the CMT (these orbits we will define as type 1 orbits), particle orbits dominated by trapping in the sides of the CMT are described in Section~\ref{type-2} (we define these as type 2 orbits, and introduce three subcategories of these in Section~\ref{frequencies}), while orbits which escape the CMT are described in Section~\ref{type-3} (we define these orbits as type 3).

%\subsection{TRAPPING IN JET BRAKING REGION (TYPE 1)}\label{type-1}
\subsection{Trapping in Jet Braking Region (Type 1)}\label{type-1}
\begin{figure}[h!]
	\subfloat[][Test particle trajectory]{\includegraphics[width = 0.47\textwidth]{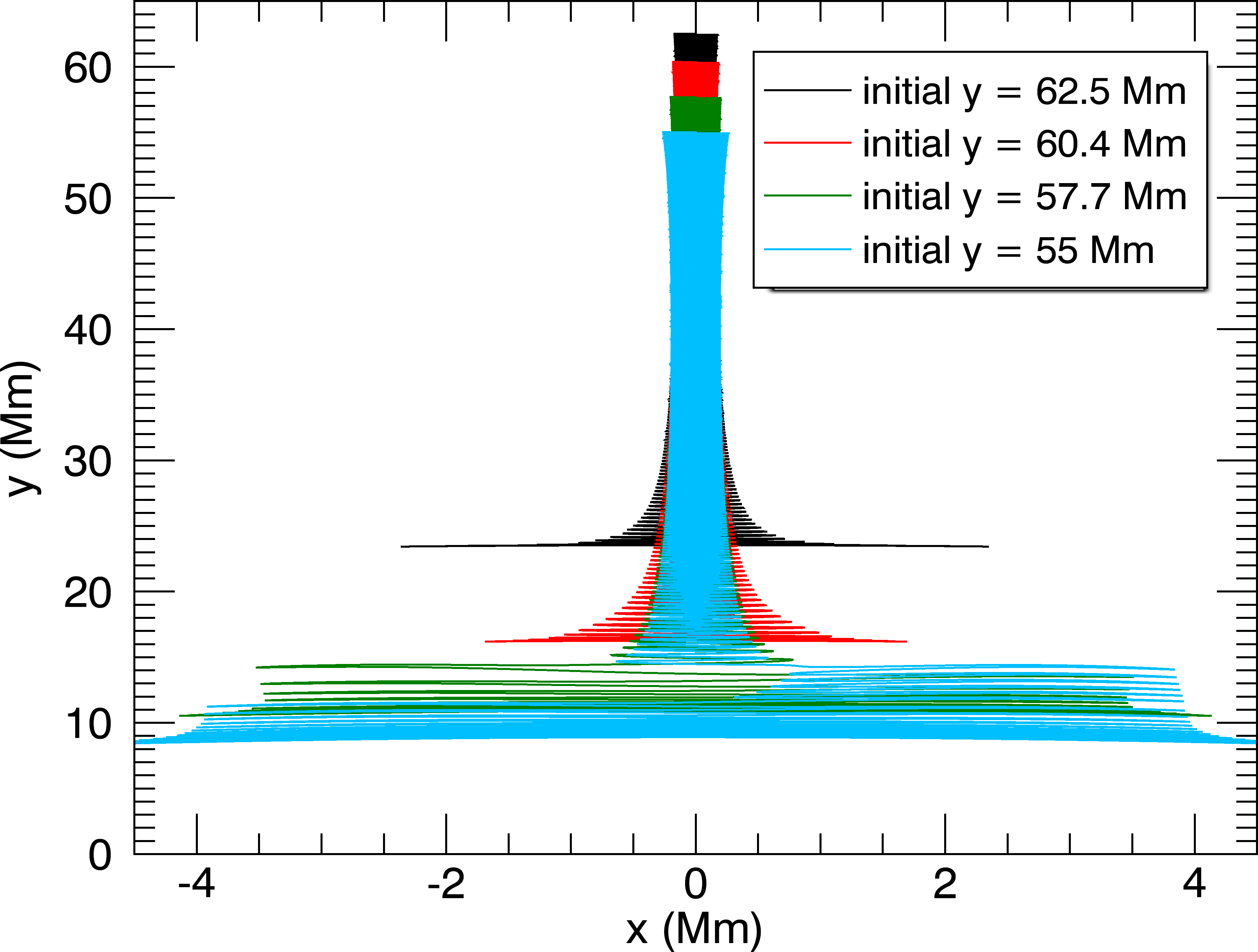}}
	\qquad
	\subfloat[][Test particle energy]{\includegraphics[width = 0.47\textwidth]{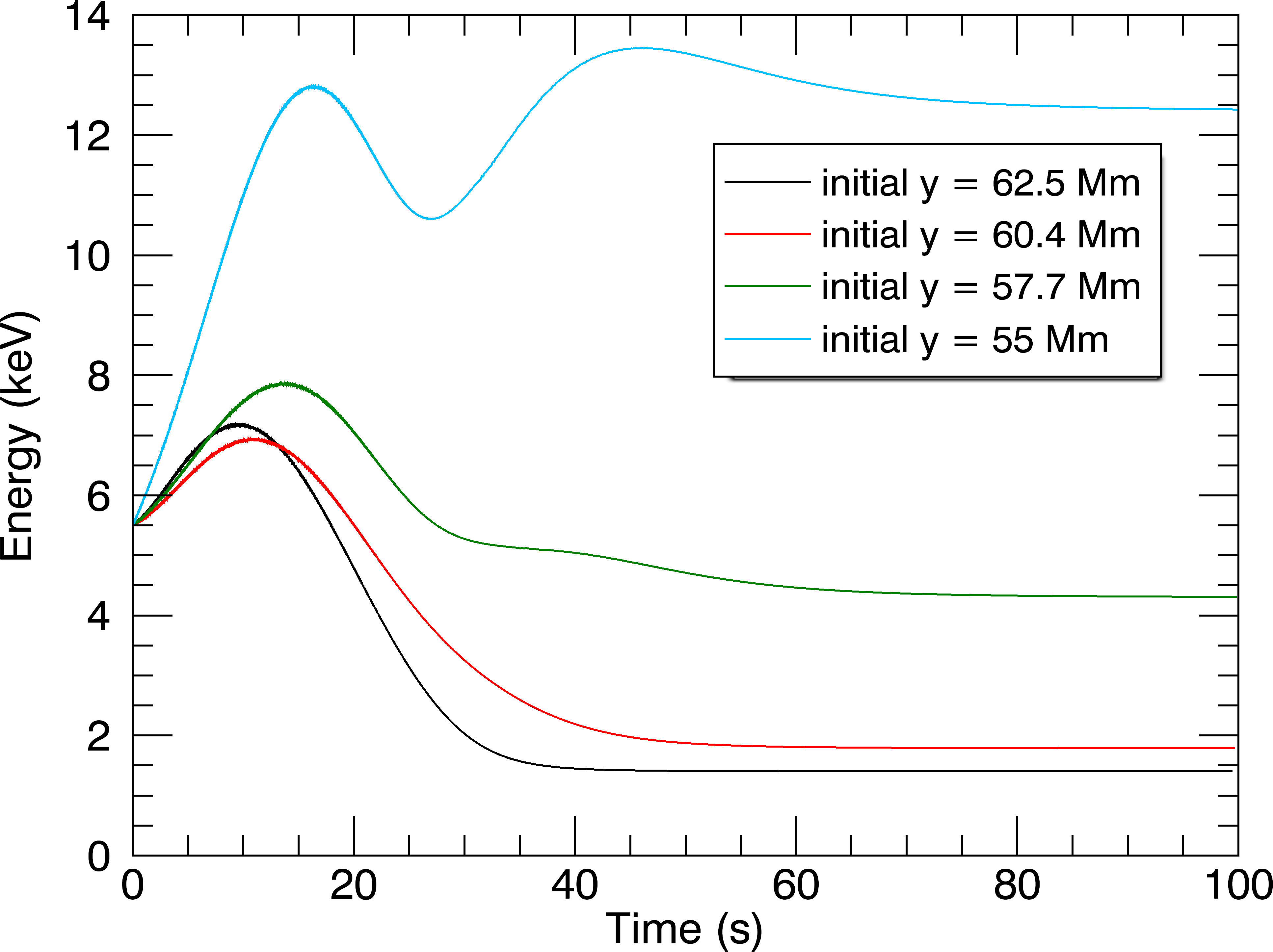}}
	\caption{Test particle trajectories (a) and kinetic energies (b) in cases where orbits become trapped in the jet braking region of a CMT (type 1 orbits). The initial conditions for these orbits are $x = 0\unit{Mm}, \theta = 75\deg, E_{k0} = 5.5\unit{keV}$. Initial vertical positions given in the legend.}
	\label{particle1}
\end{figure}

Test particle orbits starting in the centre of the trap with a high pitch angle are trapped near the centre in the jet braking region. To demonstrate this effect, four orbits are initialized at different vertical positions in the middle of the trap with a starting pitch angle of $\theta = 75\deg$. These initial conditions ensure that the orbits are trapped by the braking jet. The particle orbit trajectories and kinetic energies are shown in Figure~\ref{particle1}. Test particle orbits with a lower initial vertical position experience more acceleration and propagate further down in the magnetic trap. Towards the end of the simulation time, as the jet front weakens, the test particles exit the braking jet and become trapped in a wider region. Although, by definition type 1 orbits are dominated by trapping in the jet braking region, more complicated behaviour is also possible. One example of this is shown by the particle orbit started at $55~\unit{Mm}$, which is briefly trapped in the curved field lines beside the jet braking region before the front dissipates. For this test particle orbit, trapping in this region produces a small energy gain of approximately $3\unit{keV}$ starting at $t \approx 30\unit{s}$. Test particle orbits with higher initial vertical positions do not exhibit this behaviour and hence do not gain any energy from trapping in the loop sides. 

%\subsection{TRAPPING IN LOOP LEGS (TYPE 2)}\label{type-2}
\subsection{Trapping in Loop Legs (Type 2)}\label{type-2}
Test particle orbits with a lower initial pitch angle will not be trapped in the braking jet. For a low initial pitch angle and with initial horizontal position $x = 0~\unit{Mm}$, orbits initialized high in the trap, inside the front, escape the trap after few or no bounces. These orbits are discussed in Section~\ref{type-3}. In this section we focus on orbits that are initialized below the jet braking region. As the front passes, these orbits become trapped between a mirror point near to the footpoint of the loop and the braking jet. We refer to these orbits as trapped at the sides of the loop, or type 2 orbits. 

\begin{figure}[h!]
	\subfloat[][Test particle trajectory]{\includegraphics[width = 0.47\textwidth]{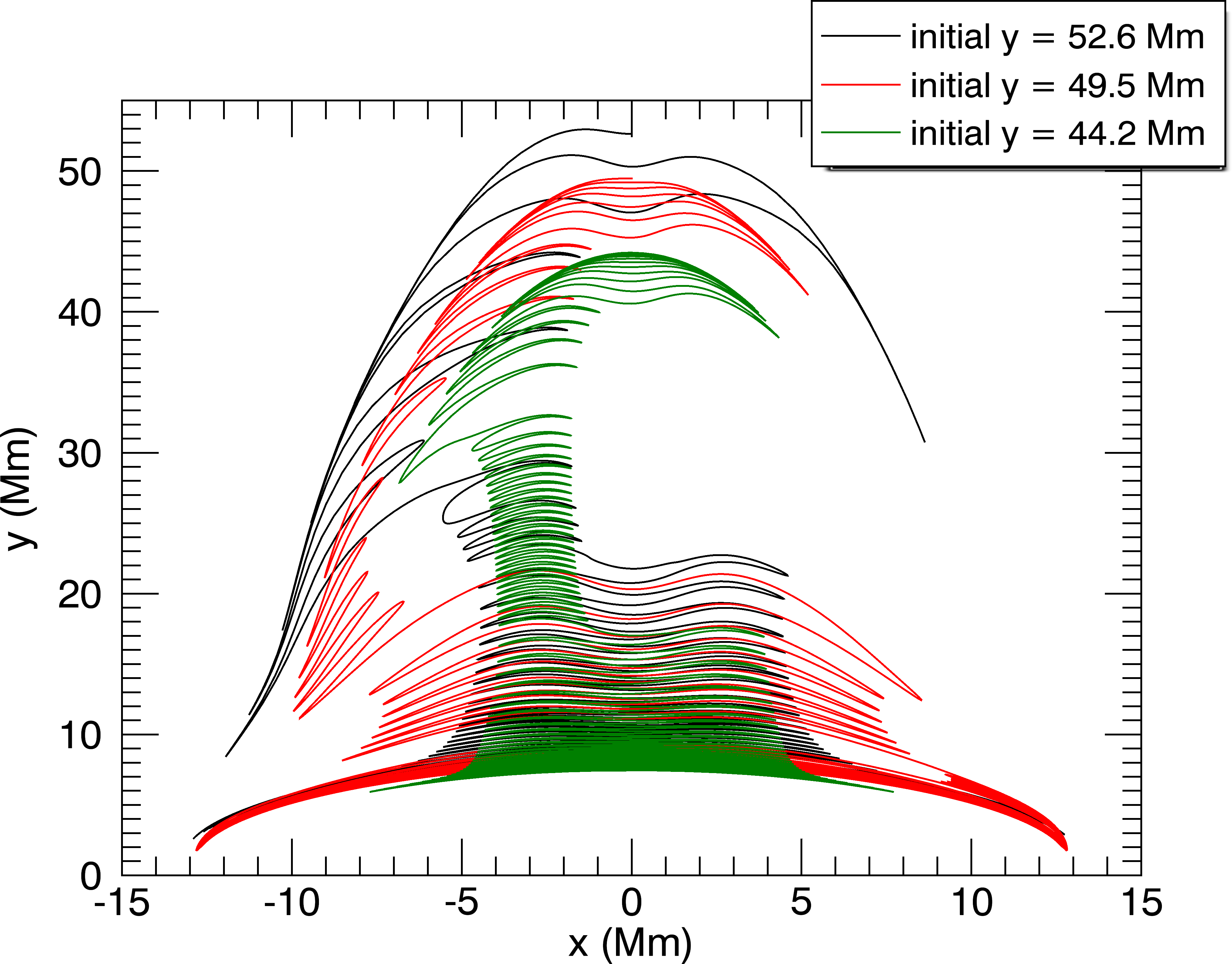}}
	\qquad
	\subfloat[][Test particle energy]{\includegraphics[width = 0.47\textwidth]{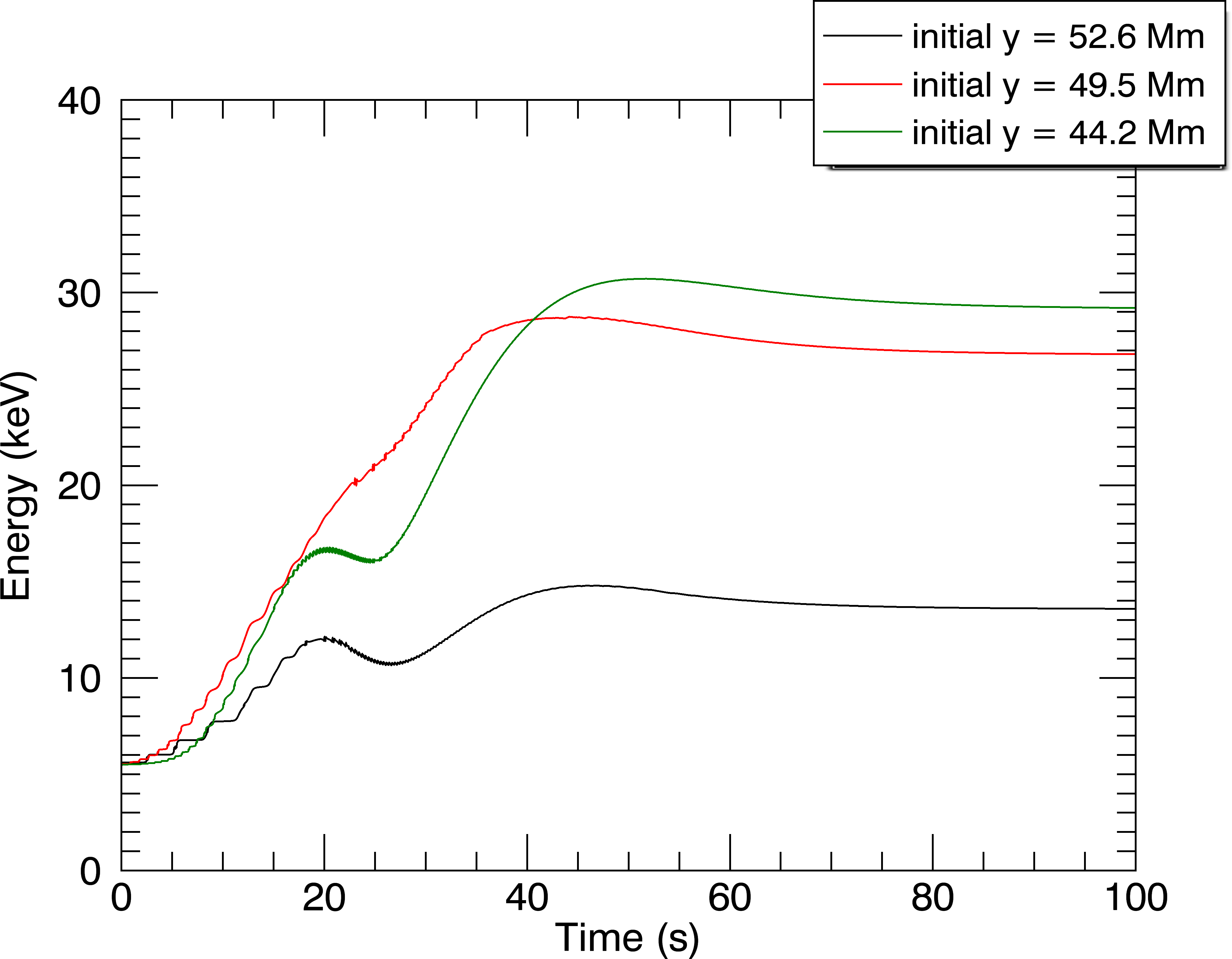}}
	\caption{Test particle trajectories (a) and kinetic energies (b) in cases where orbits become trapped in the loop legs of a CMT (type 2 orbits). The initial conditions for these orbits are $x = 0~\unit{Mm}, \theta = 40\deg, E_{k0} = 5.5~\unit{keV}$. Initial vertical positions are given in the legend.}
	\label{particle2}
\end{figure}

To demonstrate typical type 2 behaviour we again initialize four test particle orbits with varying initial vertical position at the centre of the CMT ($x = 0~\unit{Mm}$), and an initial pitch angle of $\theta = 40\deg$. The resulting particle orbits and kinetic energies are presented in Figure~\ref{particle2}. Test particles are initialized on a field line which has only one bend in it prior to the passage of the jet front. Test particles on such a field line travel between two mirror points on either side of $x = 0~\unit{Mm}$. As the jet front approaches, the field lines start to curve and compress. This compression increases the magnetic field strength at the braking jet leading to particle orbits becoming confined away from the jet centre. As the trap collapses particle orbits gain energy. Towards the end of the simulation, the field lines straighten again and the test particles no longer mirror at the jet centre, but instead have access to a larger portion of the lower loop. 

The energy gains of type 2 orbits do not follow the simple pattern of type 1 orbits. Very little energy is gained before the jet front passes the field line along which the test particle is orbiting. Once the jet reaches the test particle orbit location, the test particle experiences significant acceleration while trapped in the side of the loop. Furthermore we see that there are slightly different locations where orbits are trapped at the side of the loop. For instance, the orbit started at $49.5~\unit{Mm}$ gets trapped lower in the loop leg and as a result gains energy without interruption (in comparison to the orbit started at $44.2\unit{Mm}$ which changes locations where it is trapped more often).

\begin{figure}[h!]
	\subfloat[][Test particle trajectory]{\includegraphics[width = 0.47\textwidth]{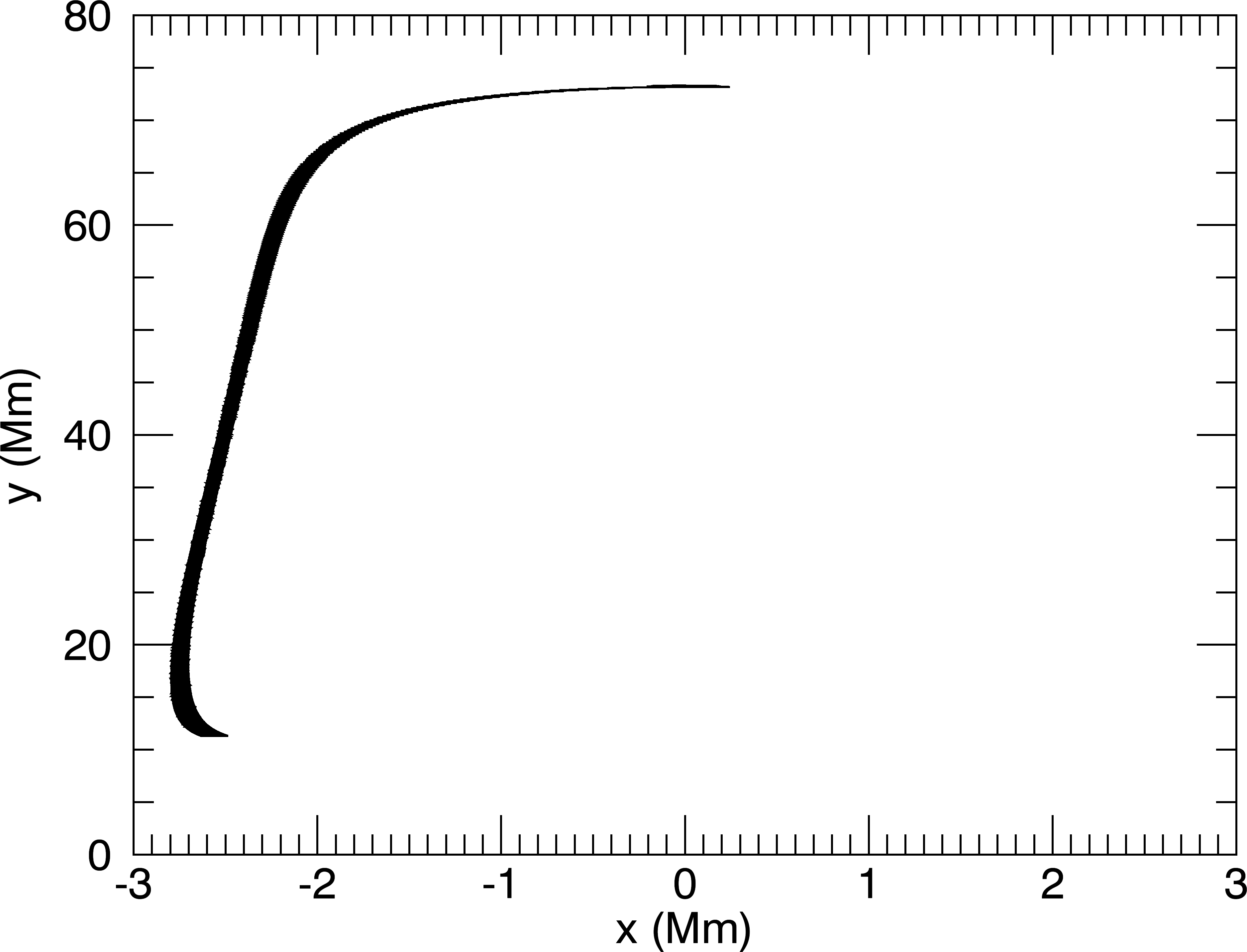}}
	\qquad
	\subfloat[][Test particle energy]{\includegraphics[width = 0.47\textwidth]{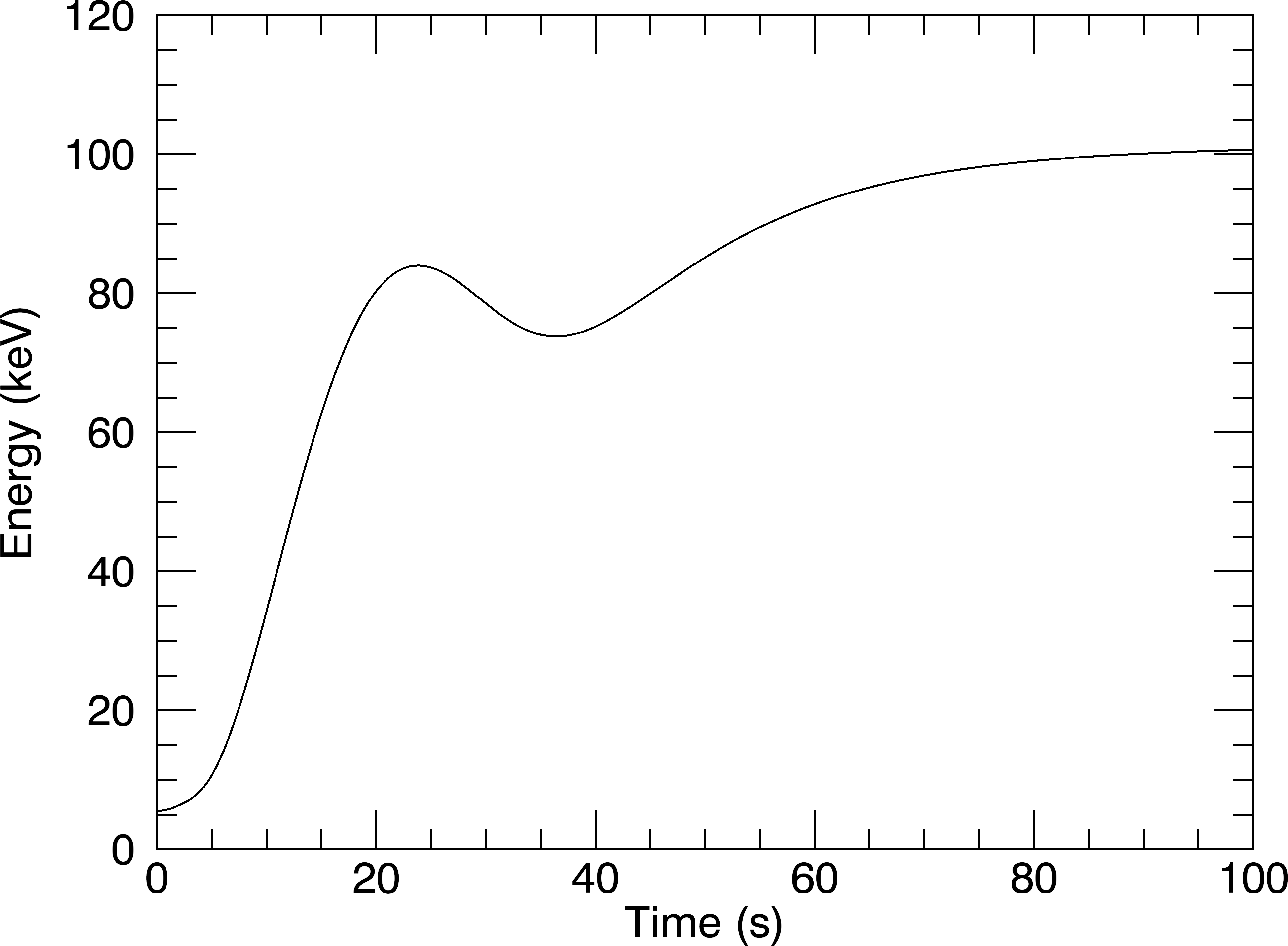}}
	\caption{Test particle orbit (a) and kinetic energy (b) for an orbit achieving kinetic energies higher than $100~\unit{keV}$. The trap parameters used in this simulation are the same as given in Table \ref{parameters} with the following parameters modified: $v_\phi = -3$, $\sigma = 3$, and $y_0 = 10$. The initial conditions of the test particle orbit are: $y = 73.3\unit{Mm}, x = 0\unit{Mm}, \theta = 85\deg, E_{k0} = 5.5\unit{keV}$.}
	\label{high-energy}
\end{figure}

Of the categories of orbits found in this CMT, type 2 motion shows the largest energy gains. By varying the trap parameters it is possible to obtain energies of $\approx 100$ keV for electrons (see Section~\ref{frequencies} for effects of trap parameters on particle motion).  An example of an orbit achieving such energies is shown in Figure~\ref{high-energy}, which is achieved by setting $v_\phi = -3$, $\sigma = 3$, and $y_0 = 10$, resulting in a faster initial jet flow speed (see the Appendix for an explanation of how these parameters affect the location and speed of the front). 

%\subsection{EARLY ESCAPE (TYPE 3)}\label{type-3}
\subsection{Early Escape (Type 3)}\label{type-3}
\begin{figure}[h!]
	\subfloat[][Test particle trajectory]{\includegraphics[width = 0.47\textwidth]{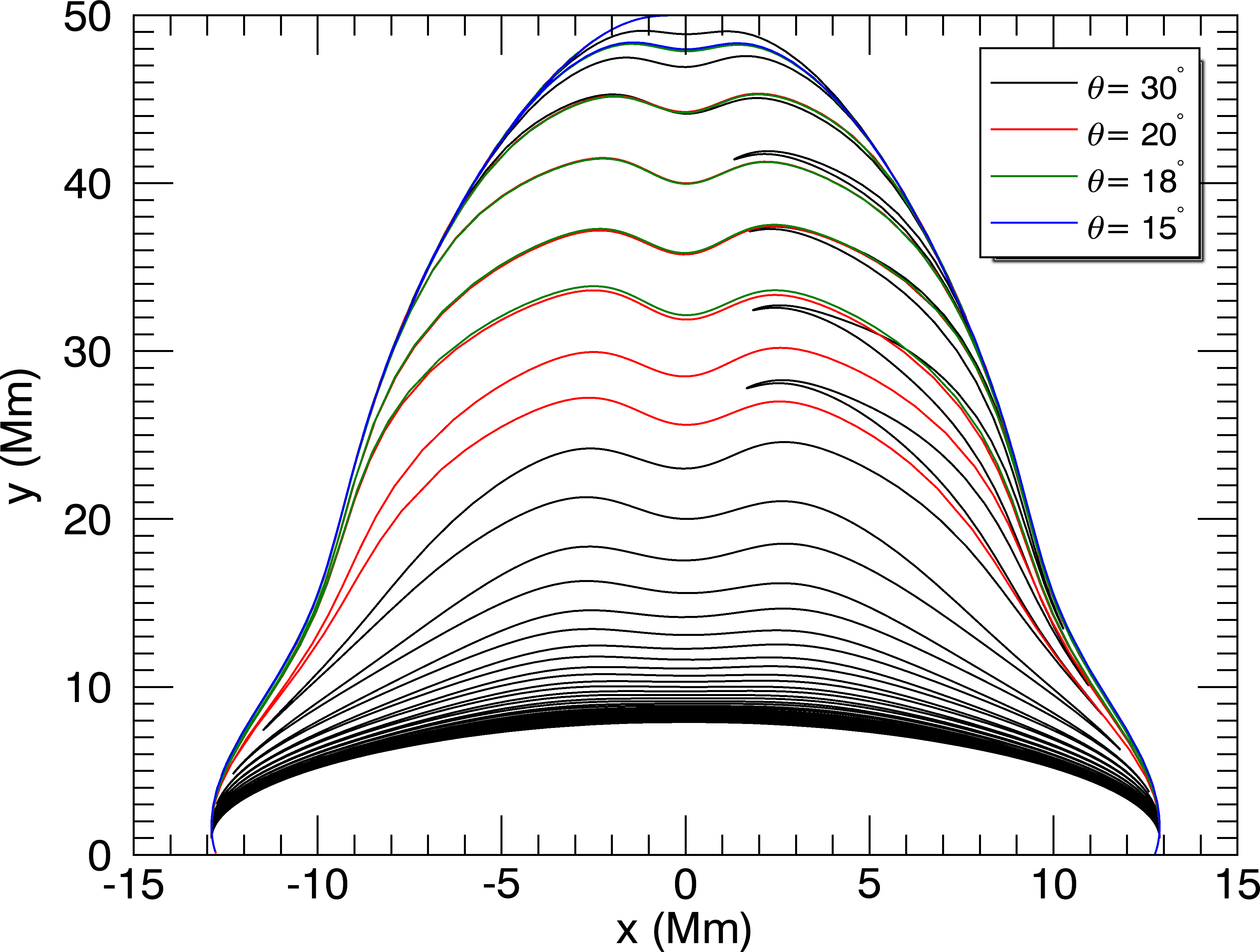}}
	\qquad
	\subfloat[][Test particle energy]{\includegraphics[width = 0.47\textwidth]{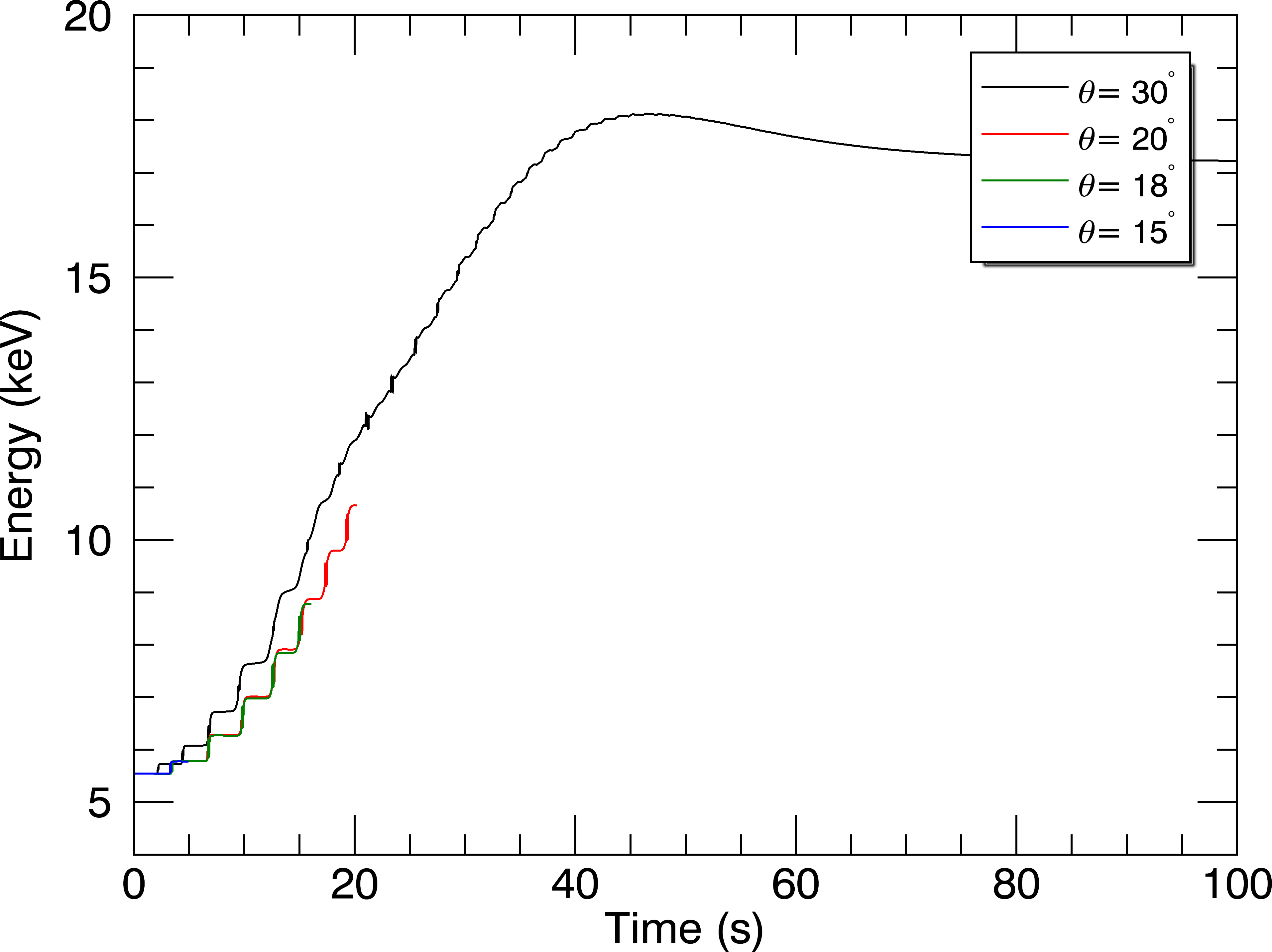}}
	\caption{Test particle trajectories (a) and kinetic energies (b) in cases where orbits exit the CMT within the simulation time (type 3 orbits), with the exception of the orbit shown in black, which is a type 2 orbit, shown for comparison. The initial conditions of the test particle orbit are: $y = 50\unit{Mm}, x = 0\unit{Mm}, E_{k0} = 5.5\unit{keV}$. The initial pitch angles are given in the legend. }
	\label{particle3}
\end{figure}

As discussed in Section~\ref{type-2}, test particle orbits initialized with a low pitch angle near the jet front are unlikely to be trapped for more than a few bounces; as a result these orbits escape much earlier and do not gain as much energy as type 2 orbits or some type 1 orbits. We refer to these as type 3 orbits. To demonstrate the dependence of the type of orbit on the initial conditions, we present four orbits starting at the same position in the CMT, with varying initial pitch angle in Figure~\ref{particle3}. Test particles with initial pitch angle $\theta = 15\deg - 20\deg$ develop type 3 orbits and all escape the CMT after about 20 seconds. By increasing the initial pitch angle, test particle orbits execute increasing numbers of bounces, until at $\theta = 30\deg$ they no longer exit the CMT within the simulation time. This is reminiscent of the findings of \citet{eradat_oskoui-et-al2014}, albeit in the context of a different CMT model. Further increases in initial pitch angle for the same initial position yield type 2 orbits and, for sufficiently high pitch angles, type 1 orbits. Among type 3 orbits, particles that are trapped longer are able to gain more energy. 

In Figure~\ref{particle3}, the particle orbit with initial pitch angle $\theta = 30\deg$ is more readily classified as a type 2 orbit because it remains trapped throughout the simulation time and for a portion of the orbit is confined to the loop leg. Test particles with a lower initial pitch angle have type 3 orbits. This shows that small changes in initial conditions can cause an orbit to entirely change classification. 

%%%%%%%%%%%%%%%%%%%%%%%%%%%%%%%%%%%%%%%%%%%%%%%%%%%%%%%%%%%%%%%%%%%%%%%%%%%%%%%%%%%%%%%%%%%%%%%%%%%%%%%%%%%%%%%%%%%%%%%%

\section{Discussion of Test Particle Motion and Acceleration Mechanisms}\label{particle-motion}

%\subsection{TRAPPING LOCATIONS}
\subsection{Trapping Locations}

\begin{figure}
\centering
	\includegraphics[width = 0.7\textwidth]{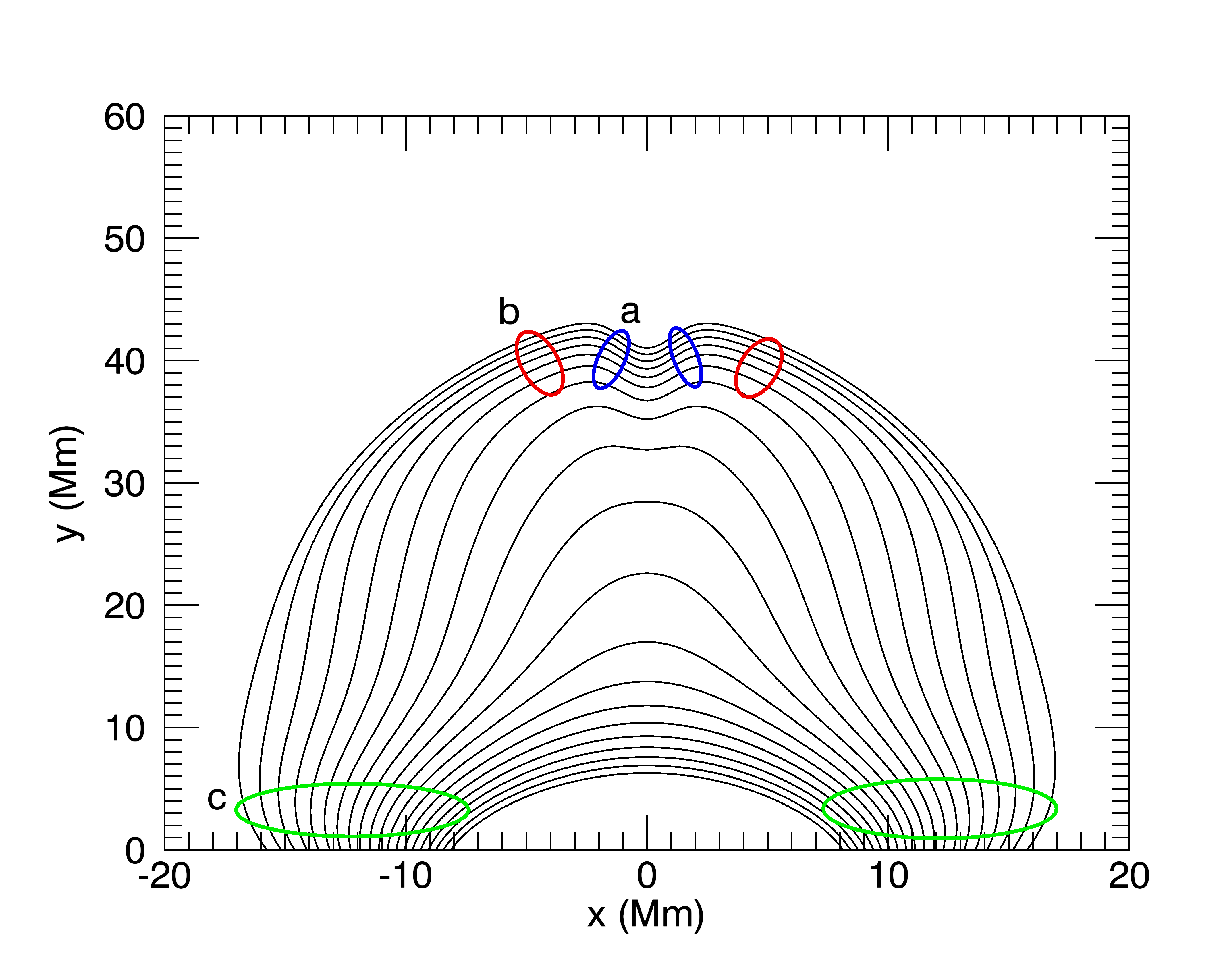}
	\caption{Approximate locations of possible mirror points for particles trapped in our CMT model are circled. The test particle orbit initial conditions determine in which region the particle is mirrored. Black lines are magnetic field lines. Regions a and b may contain mirror points because of the higher magnetic field strength associated with the braking jet. Mirror points located in the region c occur due to the stronger magnetic field at the solar surface.}
	\label{mirror}
\end{figure}

The behaviour of particle orbits outlined in Section~\ref{sample-trajectories} may be explained by considering the positions of possible mirror points within the CMT. Mirroring of a particle occurs due to terms on the right hand side of Equation~(\ref{NRGC2}). In our model mirror points can occur due to field line curvature (first term on the right hand side of Equation~(\ref{NRGC2})) and strengthening of the magnetic field (third term on the right hand side of Equation~(\ref{NRGC2})\footnote{There are no contributions to the parallel guiding centre velocity from the parallel electric field since $E_\pr = 0$ in ideal MHD. This means that mirroring due to the parallel electric field cannot occur \citep[see][]{threlfall-et-al2016}.}). For a given test particle orbit, mirror points may be located on either side of the indentation caused by the braking jet (regions labelled a in Figure~\ref{mirror}), further up the loop legs just below the front (regions labelled b in Figure~\ref{mirror}) as well as near the footpoints of the loops (regions labelled c in Figure~\ref{mirror}).  

Type 1 orbits are trapped between two mirror points located on either side of the indentation around the region of the braking jet (labelled a in Figure~\ref{mirror}). Type 2 motion is caused by mirroring between points in regions c and b, c and a, or b and a. For example, trapping between mirror points in regions b and c occurs for the orbit started at $49.5\unit{Mm}$ in Figure~\ref{particle2}, whereas the test particle orbit initialized at $44.2\unit{Mm}$ in Figure~\ref{particle2} mirrors between points in regions a and b. As the jet progresses downward, the magnetic field strength increases, causing some particle orbits to be trapped outside of the jet braking region, resulting in type 2 motion. In order to see an appreciable front in this CMT model the magnetic field strength in the front needs to be comparable to the magnetic field near the footpoints. As a result, test particle orbits with a sufficiently small initial pitch angle to escape the jet braking region gain enough energy within a few bounces to escape the trap.

%\subsection{ACCELERATION MECHANISMS}
\subsection{Acceleration Mechanisms}

\begin{figure}
	\subfloat[][Type 1 orbit]{\includegraphics[width = 0.47\textwidth]{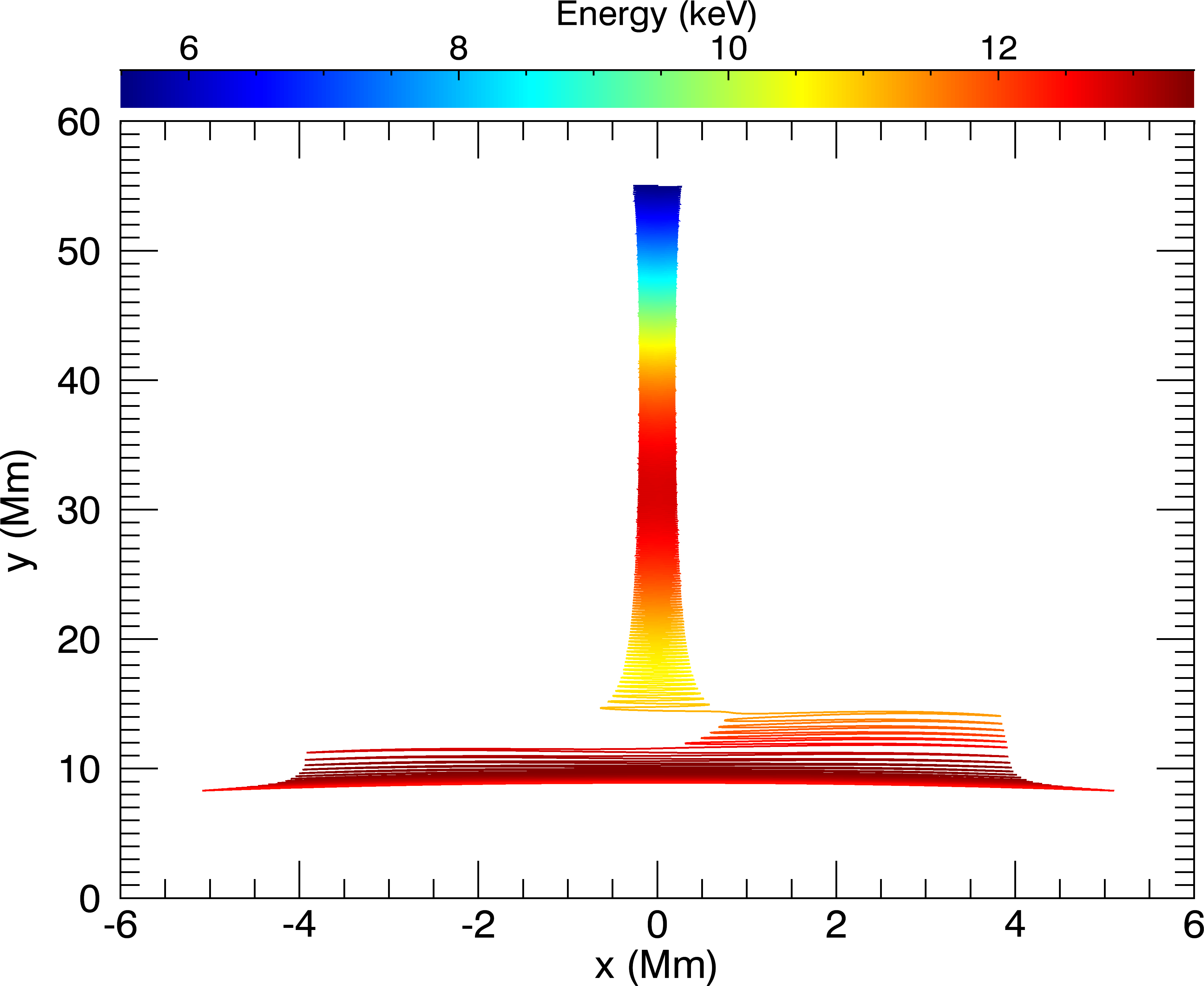}}
	\qquad
	\subfloat[][Type 2 orbit]{\includegraphics[width = 0.47\textwidth]{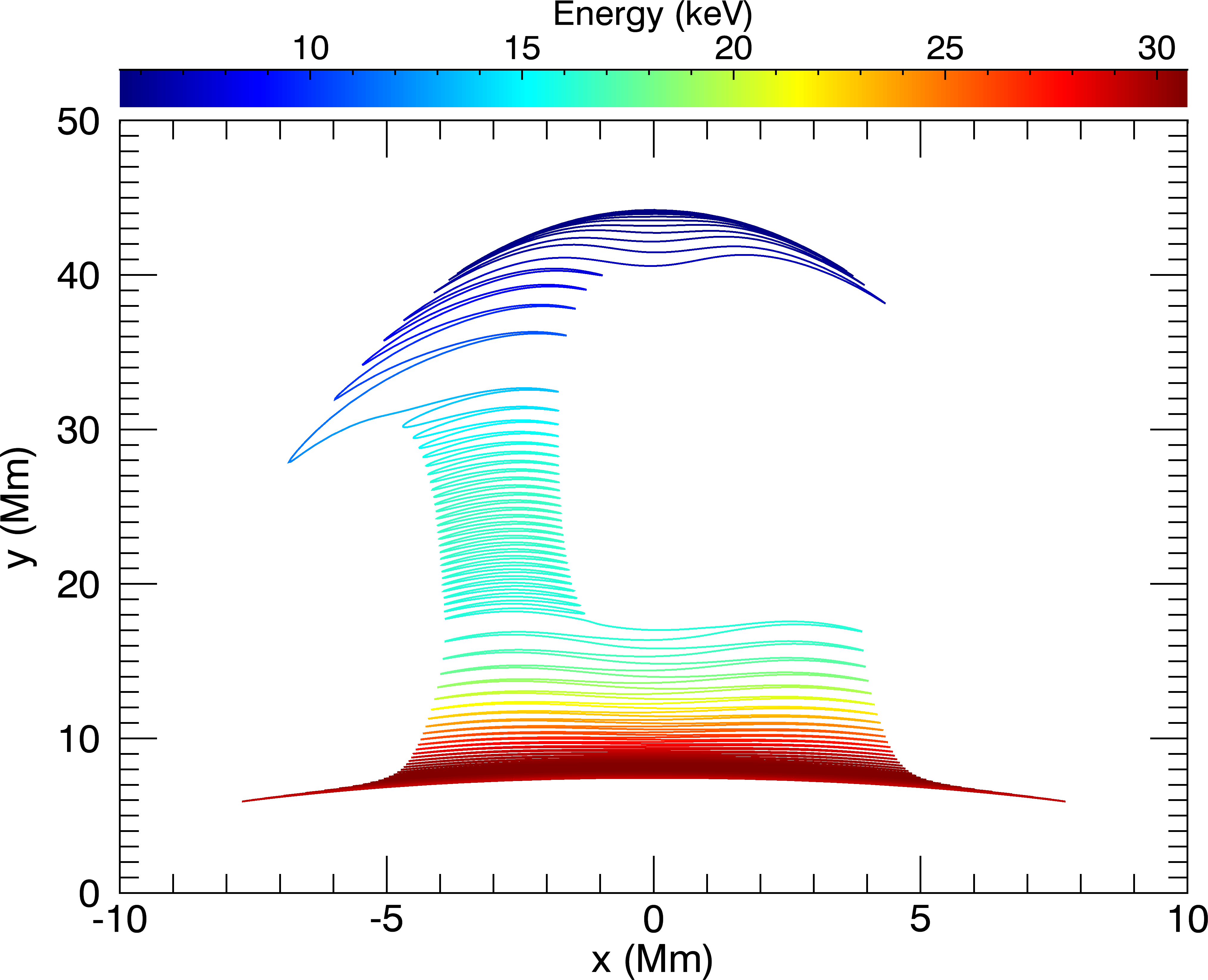}}
	\caption{Particle orbit energy, indicated in colour, with respect to position for type 1 and 2 orbits. The type 1 and 2 orbits correspond to orbits started at 55~\unit{Mm} in Figure~\ref{particle1} and at 44.2~\unit{Mm} in Figure~\ref{particle2} respectively. The type 1 orbit shows an initial increase in energy, followed by a decrease while trapped in the braking jet. When the orbit escapes the jet and becomes trapped in the side of the loop there is an associated increase in energy. The type 2 orbit shows sharp increases in energy for $40\unit{Mm} \ge y \ge 30~\unit{Mm}$ and for $15\unit{Mm} \ge y \ge 9\unit{Mm}$.}
	\label{energy-evol}
\end{figure}

To ascertain the location of most efficient acceleration for a particular orbit we consider Figure~\ref{energy-evol}, which shows how the energy of two test particles (of type 1 and 2) changes throughout their respective orbits. The type 1 orbit (Figure~\ref{energy-evol}a) corresponds to the orbit started at 55~\unit{Mm} in Figure~\ref{particle1}. This orbit demonstrates the typical motion of type 1 particles when trapped in the braking jet, in addition to showing some trapping in the loop sides lower in the orbit. Type 1 orbits experience an initial increase, followed by a decrease of energy when trapped in the braking jet. This is caused in part by initially increasing magnetic field strength while the jet is propagating through regions of low background field, followed by decreasing magnetic field strength when the front encounters regions of stronger field in the lower loops causing the front to dissipate. If the particle orbit exits the braking jet before the jet dissipates then the particle orbit can gain energy in the loop sides, as shown in Figure~\ref{energy-evol}a.

The type 2 orbit (Figure~\ref{energy-evol}b) corresponds to the orbit started at 44.2~\unit{Mm} in Figure~\ref{particle2}. In this case most of the energy is gained in two regions of the particle orbit. The first region occurs when the particle is confined to the side of the loop for $40\unit{Mm} \ge y \ge 30~\unit{Mm}$, while the second occurs when it is trapped lower down for $15\unit{Mm} \ge y \ge 9\unit{Mm}$ (although less dramatic increases are also present at other times when trapped in the loop sides). Similar to the type 1 orbit, the energy gains are partially caused by increases of the magnetic field, however there is no substantial decrease in magnetic field strength when the front dissipates due to the particle being trapped outside of the braking jet. 

\begin{figure}
	\subfloat[][Type 1 orbit]{\includegraphics[width = 0.47\textwidth]{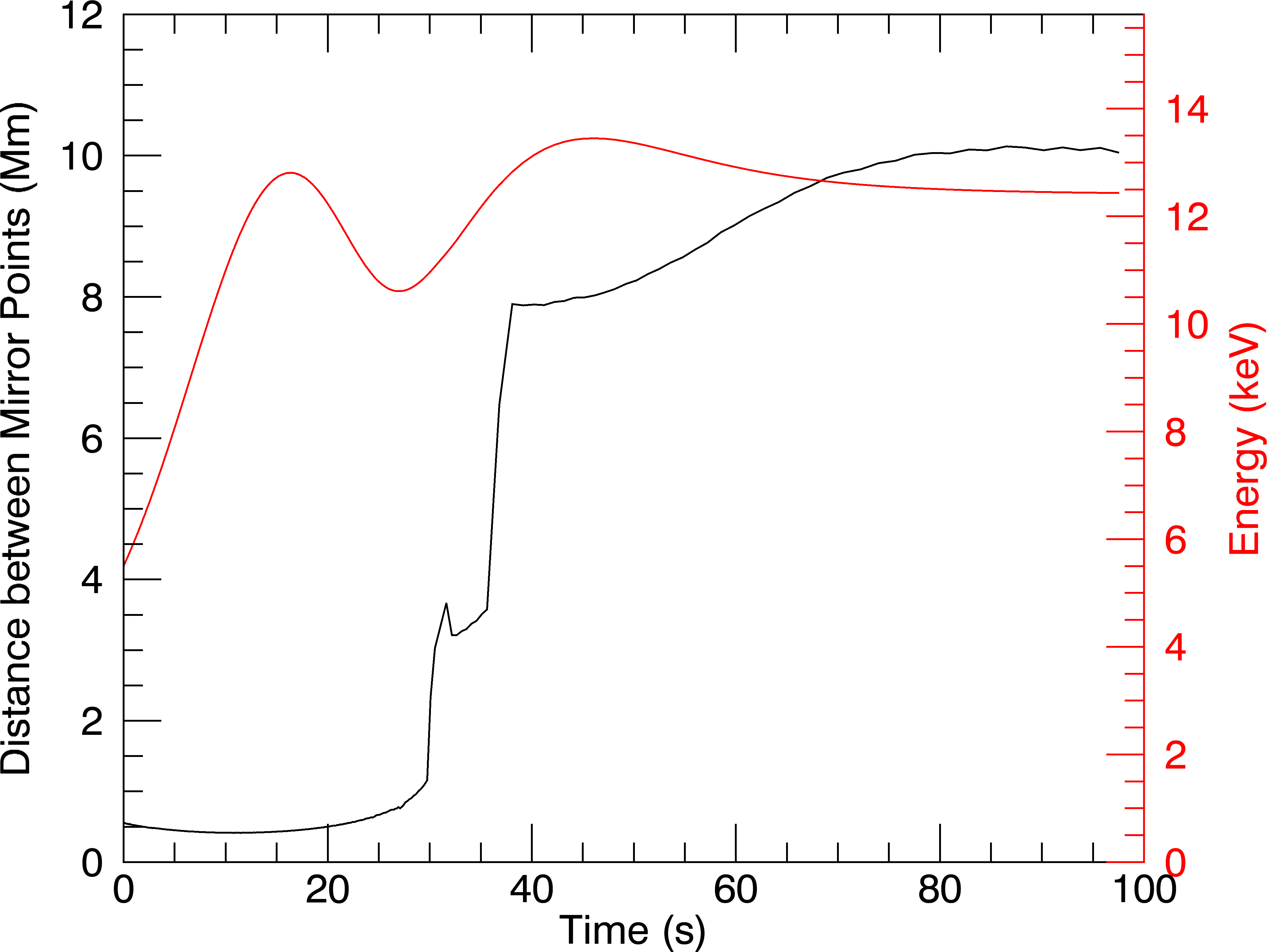}}
	\qquad
	\subfloat[][Type 2 orbit]{\includegraphics[width = 0.47\textwidth]{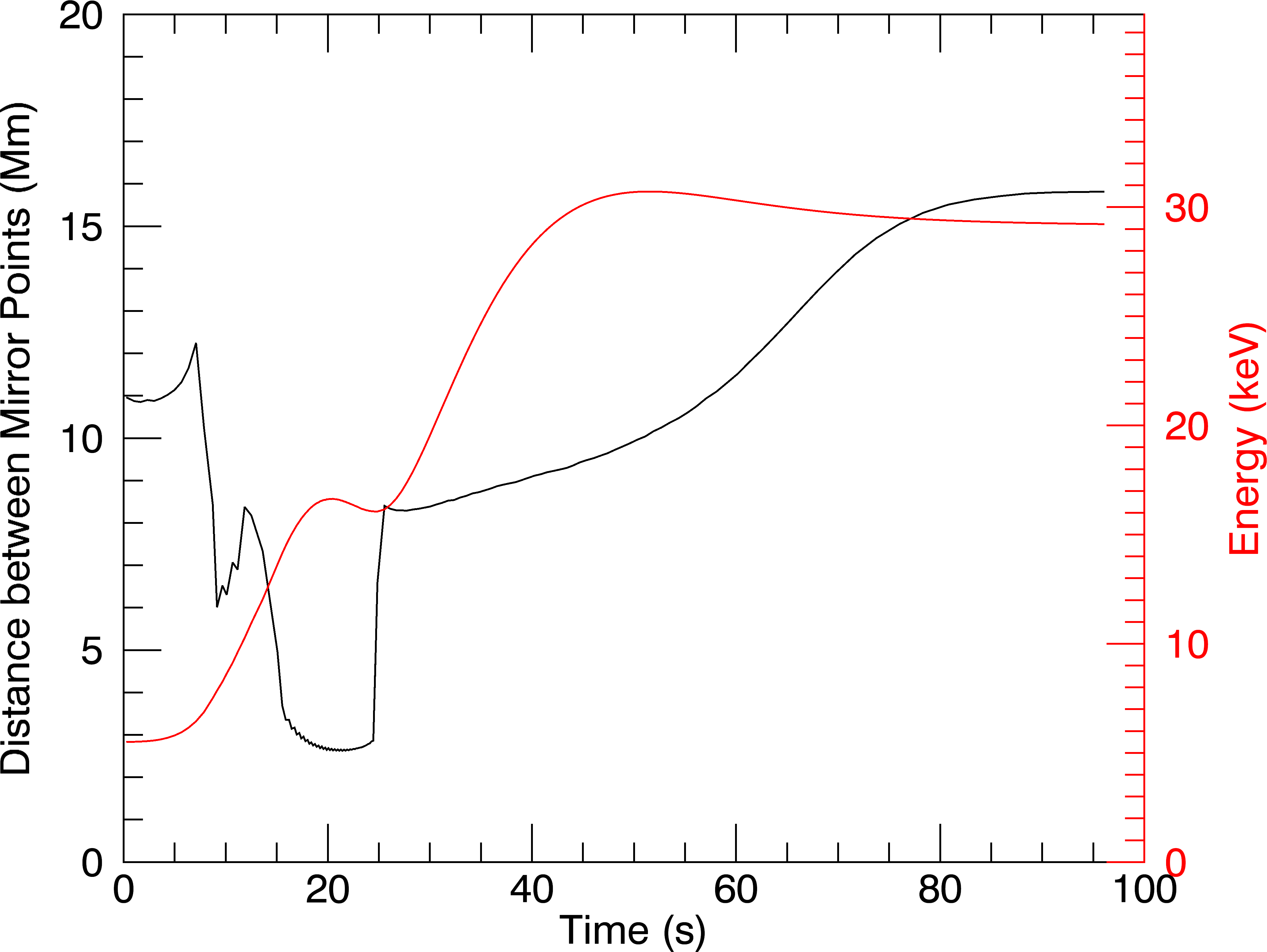}}
	\caption{Distance between bounces (black) and kinetic energy (red) with respect to time for type 1 and 2 particle orbits. The type 1 and 2 orbits correspond to orbits started at 55~\unit{Mm} in Figure\ref{particle1} and at 44.2~\unit{Mm} in Figure~\ref{particle2} respectively. We see that the distance between bounces decreases in the early stages of the type 1 orbit (for $t < 10\unit{s}$) and for a significant portion of the type 2 orbit ($t < 25\unit{s}$), which may be indicative of Fermi acceleration affecting the energy in addition to the betatron effect.}
	\label{bounces}
\end{figure}

Despite the importance of the magnetic field strength in determining the energy of the test particle orbit, Fermi acceleration can also play a role. In Figure~\ref{bounces} the distance travelled by the particle orbit along a field line between mirror points is compared to the energy for type 1 and 2 particle orbits. The bounce distance of the type 1 orbit decreases for $t < 10~\unit{s}$ during which time the orbit energy increases, indicating possible Fermi acceleration. The same is true for the type 2 orbit for $t < 25\unit{s}$, indicating Fermi acceleration may be important in this case also. Increases in the distance between mirror points for $t > 40~\unit{s}$ for both orbits may contribute to the energy loss in these time periods. 

As a result of the complicated nature of our model, it may be difficult to directly attribute energy gains or losses to a single mechanism. Significant changes in distance between mirror points (indicative of Fermi acceleration) may be accompanied by changing magnetic field strength due to the orbit entering a different region of the CMT. This results in energy changes that are difficult to attribute to either Fermi acceleration or the betatron effect alone. 

%%%%%%%%%%%%%%%%%%%%%%%%%%%%%%%%%%%%%%%%%%%%%%%%%%%%%%%%%%%%%%%%%%%%%%%%%%%%%%%%%%%%%%%%%%%%%%%%%%%%%%%%%%%%%%%%%%%%%%%%

\section{Effect of Varying Parameters on Particle Orbits}\label{frequencies}
To analyze the effect of changing parameters in the CMT model on the types of particle orbits, in this section we investigate the connections between the orbit classification and the average position of the test particle orbit. The average position is determined by the integral 
\begin{equation}\label{xbar}
\overline{\bb x} = \frac{1}{t_{\rm{final}}}\int_0^{t_{\rm{final}}} \bb x dt,
\end{equation}
where $t_{\rm{final}}$ is the time of the test particle's escape from the trap, or the simulation time (if the orbit remains trapped). In this section $\overline x$ and $\overline y$ refer to the average position in the $x$ and $y$ directions respectively, while $\overline {\bb x} = (\overline x, \overline y)$. Type 1 orbits are symmetric about $x = 0\unit{Mm}$ for most of their duration, which restricts the average $x$-position to $\overline x \approx 0$\unit{Mm}. This is not the case for type 2 orbits, which spend significant amounts of time in the loop legs, slightly displaced from the centre of the trap resulting in $|\overline x| > 0\unit{Mm}$. Type 3 orbits (particularly those which exit the trap after few or no bounces) have $|\overline{\bb x}| \gg 0\unit{Mm}$. These orbits are also identifiable by their small maximum energies. It is important to note that $\overline{\bb x}$ is a proxy for determining the general behaviour of particle orbits and does not correspond to the location that the orbits spend most of their time at.

\begin{figure}
	\subfloat[][Histogram of average $x$ position.]{\includegraphics[trim = 0cm 0cm 0cm 0cm,width = 0.45\textwidth]{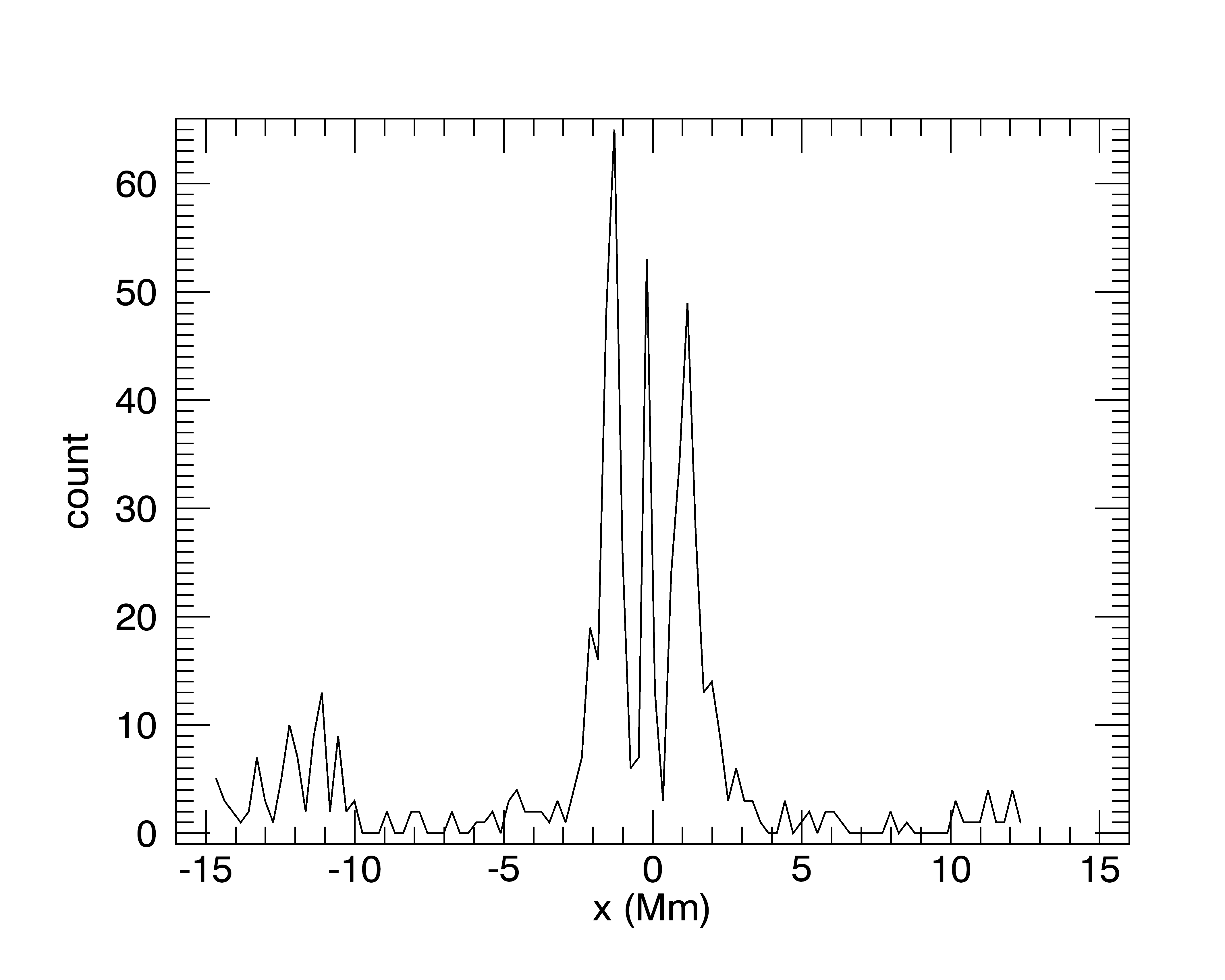}}
	\qquad
	\subfloat[][Average position.]{\includegraphics[trim = 0cm 0cm 0cm 0cm,width = 0.45\textwidth]{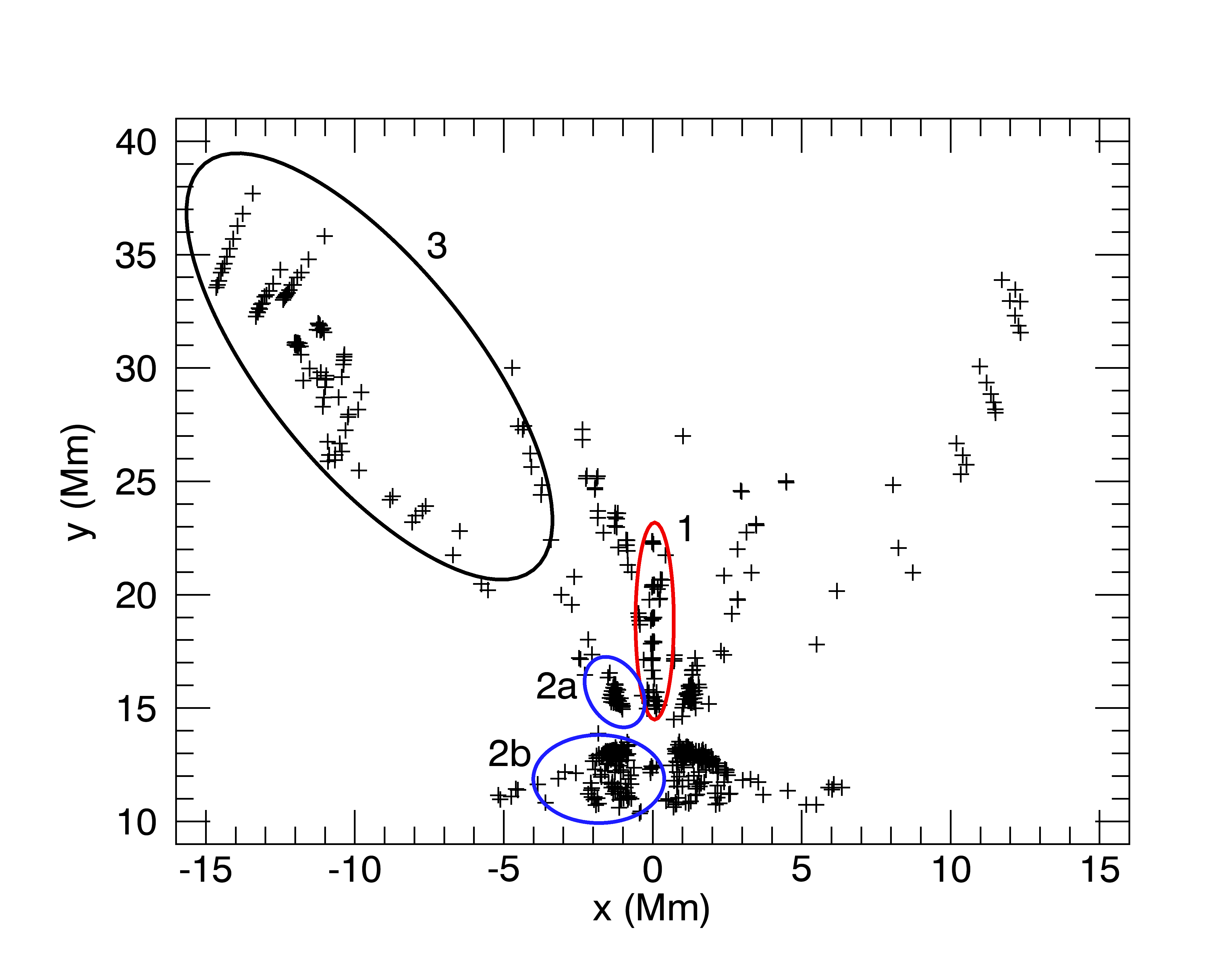}}
	\caption{Histogram of average $x$ position (a) and the average coordinate $\overline{\bb x} = (\overline x, \overline y)$ (b) for 600 particle orbits computed with the initial conditions IC1 given in Table \ref{IC} for the trap parameters given in Table \ref{parameters}. The distinction between type 1 and type 2 orbits is clearly visible in panel a, with type 2 orbits having an $|\overline x |$ between 1 and 5 Mm, while type 1 orbits have an $|\overline x |$ less than 1 Mm. Panel b shows the regions different orbit classifications occupy when their average coordinates are computed.}
	\label{explanation}
\end{figure}

The initial conditions used in the investigation are presented in Table \ref{IC}. These values are chosen to represent a meaningful part of the parameter space for the initial conditions, and to also return behaviour of interest. These values are not intended to be an accurate description of the distribution of particles in the coronal plasma during a solar flare, but are meant to illustrate how changes in trap parameters influence orbit behaviour. The results are shown in Figure~\ref{explanation}. In Figure~\ref{explanation}a, type 1 orbits are clearly identifiable as the peak near $x = 0\unit{Mm}$. The two peaks on either side of $x = 0\unit{Mm}$ correspond to type 2 orbits.

\begin{table}
\begin{tabular}{c c c c c}
\hline
Label	& x & y & $\theta$ & $E_{k0}$	\\
\hline
IC1	& 0, 2~\unit{Mm} & [45,60]~\unit{Mm} & [$25\deg,75\deg$] & 5.5~\unit{keV}	\\
IC2	& 0, 2~\unit{Mm} & [60,90]~\unit{Mm} & [$25\deg,75\deg$] & 5.5~\unit{keV}	\\
IC3	& 0, 2~\unit{Mm} & [40,55]~\unit{Mm} & [$25\deg,75\deg$] & 5.5~\unit{keV}	\\
\hline
\end{tabular}
\caption{Initial conditions used in different runs for investigating the effects of varying trap parameters on particle orbit behaviour. In each case $x$ takes on two values, $y$ takes on 15 evenly spaced values in the indicated range, and $\theta$ takes on 20 evenly spaced values in the indicated range. Initial conditions IC1 are used in conjunction with the run using basic trap parameters (see Table \ref{parameters}) and the run with a steeper front. Conditions IC2 are used in the case of a faster initial jet, while IC3 are used with a larger braking jet.}
\label{IC}
\end{table}

\begin{figure}[h!]
	\includegraphics[trim = 0cm 3cm 0cm 0cm,clip = true,width = \textwidth]{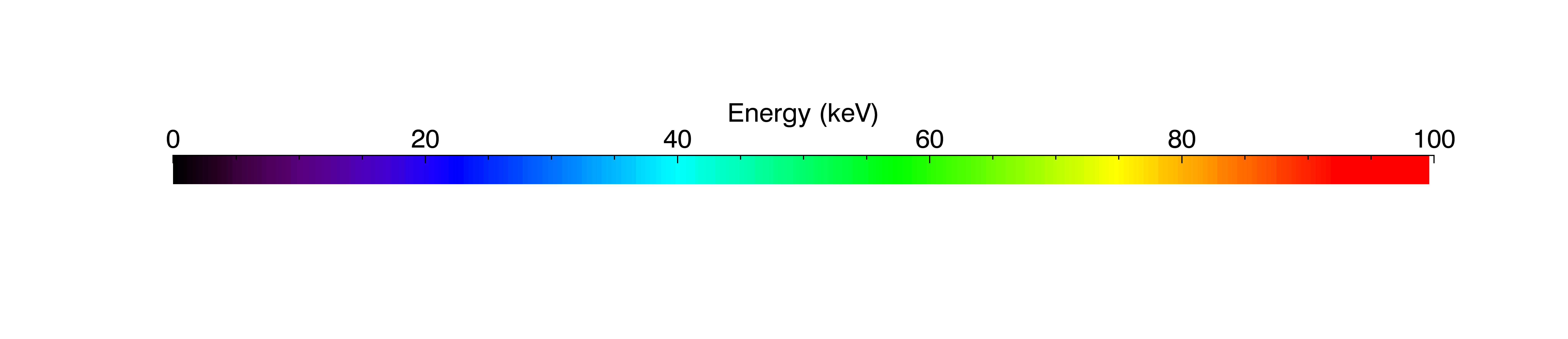}
	\\
	\subfloat[][Basic trap parameters.]{\includegraphics[width = 0.47\textwidth]{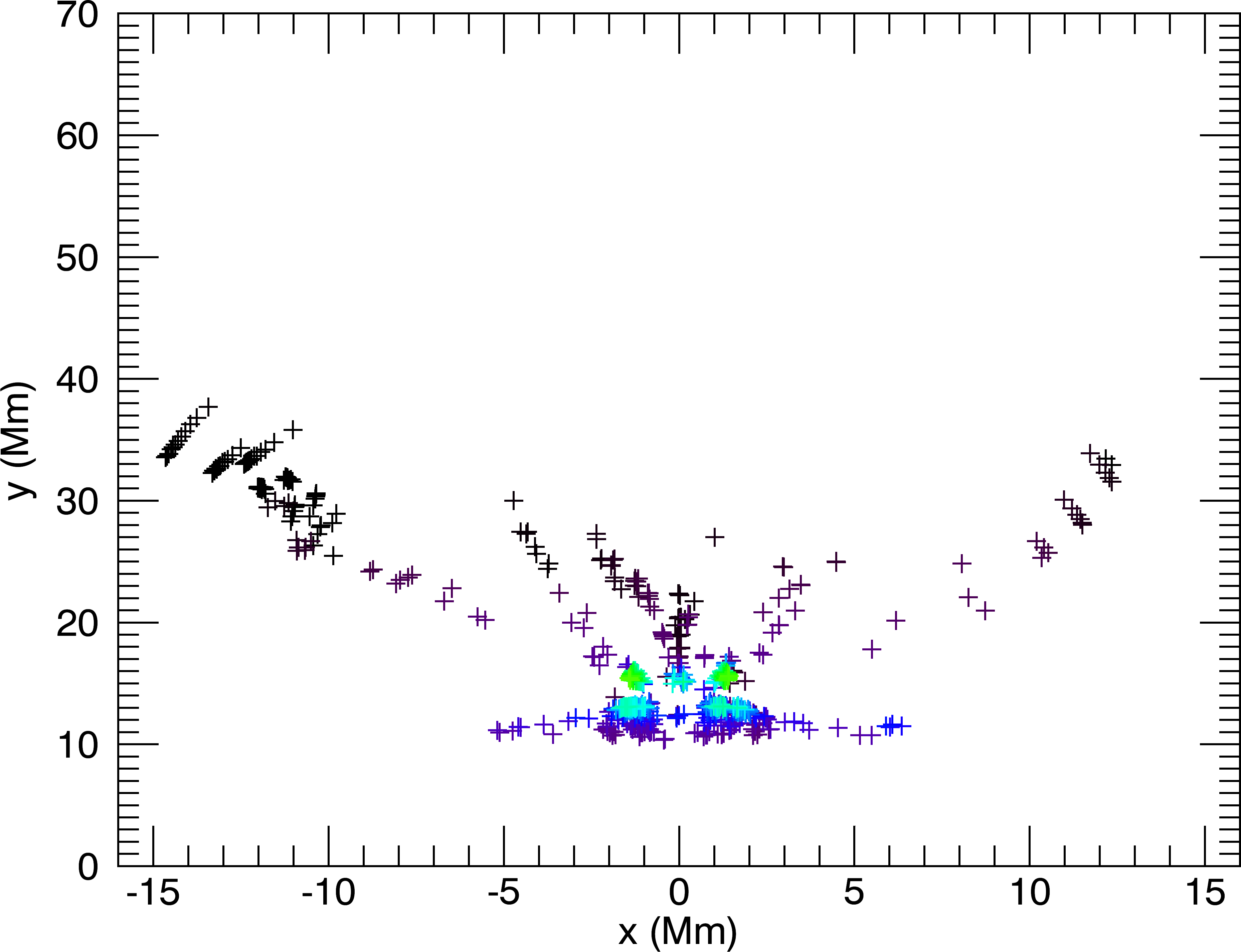}}
	\qquad
	\subfloat[][Faster initial front velocity.]{\includegraphics[width = 0.47\textwidth]{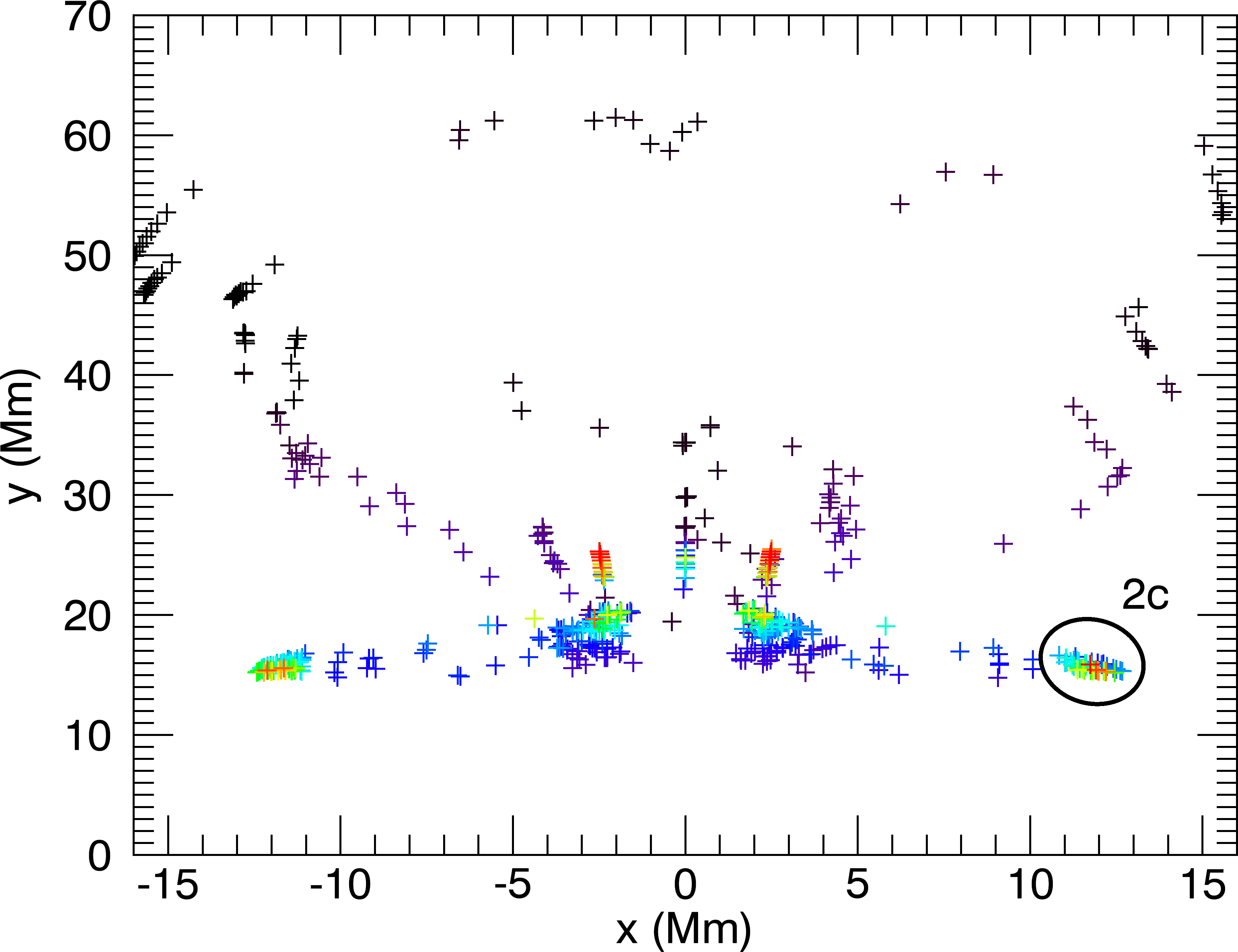}}
	\\
	\subfloat[][Larger jet braking region.]{\includegraphics[width = 0.47\textwidth]{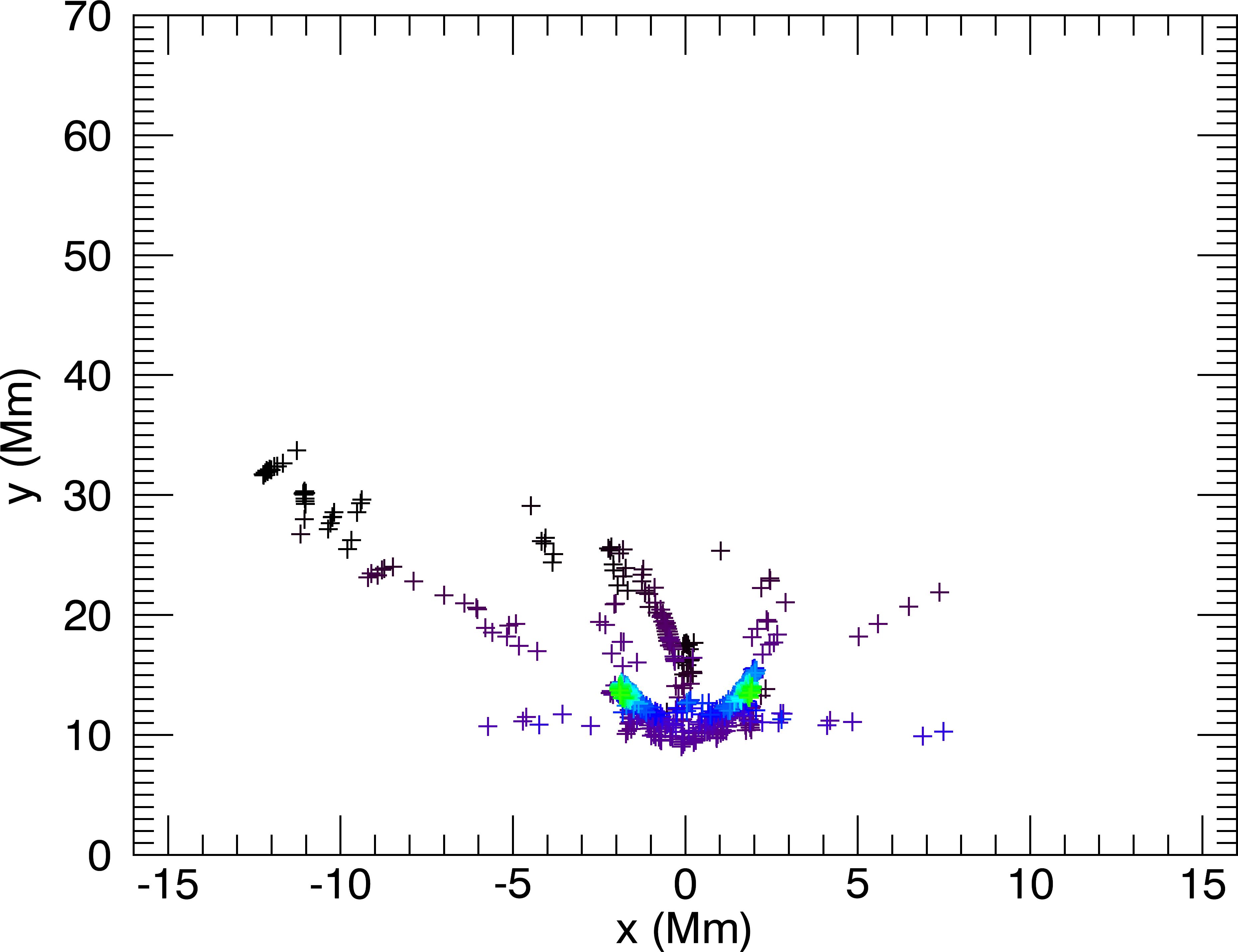}}
	\qquad
	\subfloat[][Steeper front.]{\includegraphics[width = 0.47\textwidth]{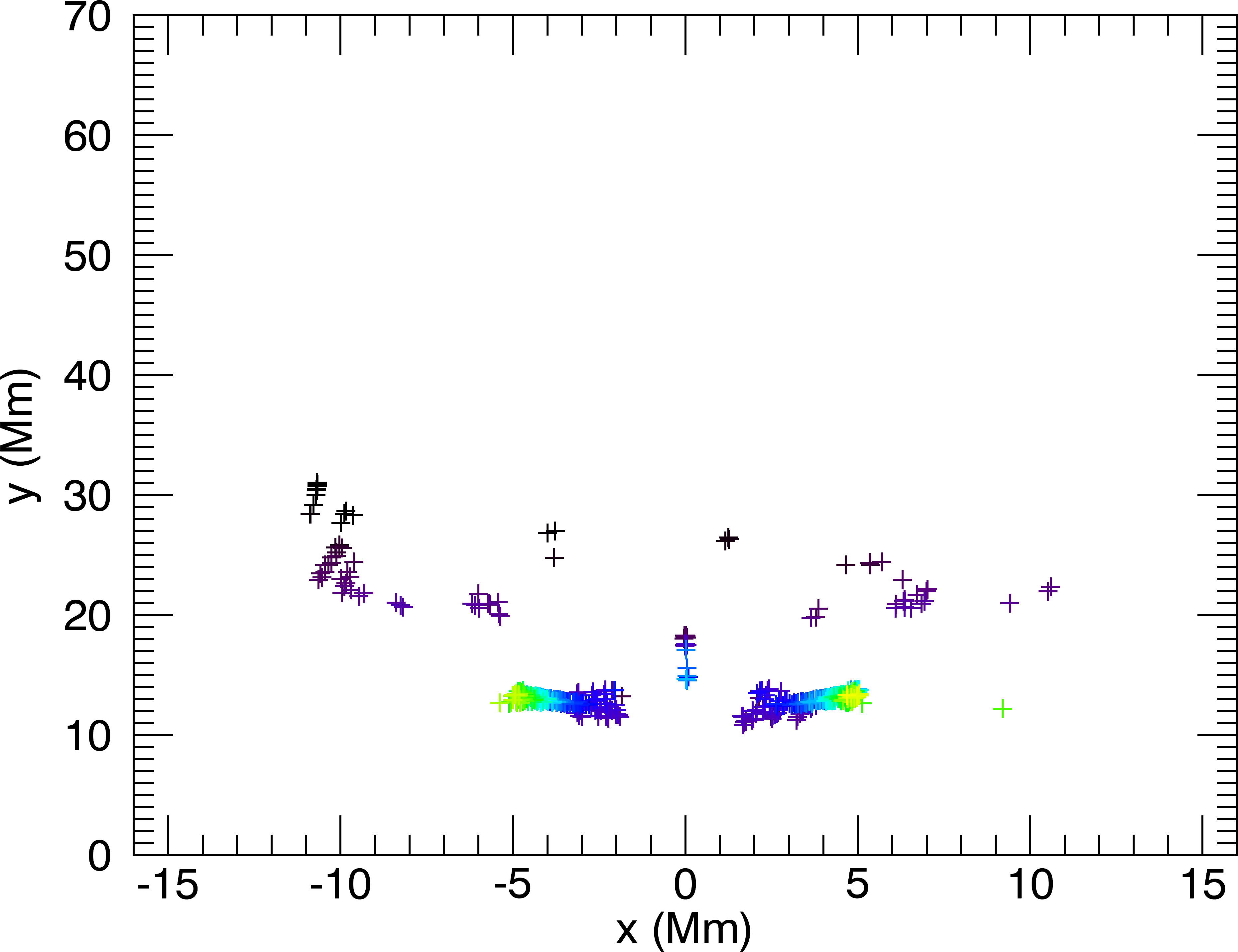}}
	\caption{Average position and energy of 600 particle orbits for various trap parameters. Trap parameters and test particle orbit initial conditions are given in Table \ref{param}.}
	\label{dataset}
\end{figure}

Figure~\ref{explanation}b shows that type 2 motion can be divided into two different categories whose average positions are fairly distinct. We label these subcategories 2a and 2b for decreasing $\overline y$ respectively. For certain choices of trap parameters these sub-categories may disappear (see Figure~\ref{dataset}c) or another category may appear (see Figure~\ref{dataset}b), which we will call 2c. Type 2c particles have average positions satisfying $\overline y < 20\unit{Mm}$ and $|\overline x| > 13\unit{Mm}$. For parameters and initial conditions for each of the following runs see Table \ref{param}.

\begin{figure}[h!]
	\includegraphics[width = \textwidth]{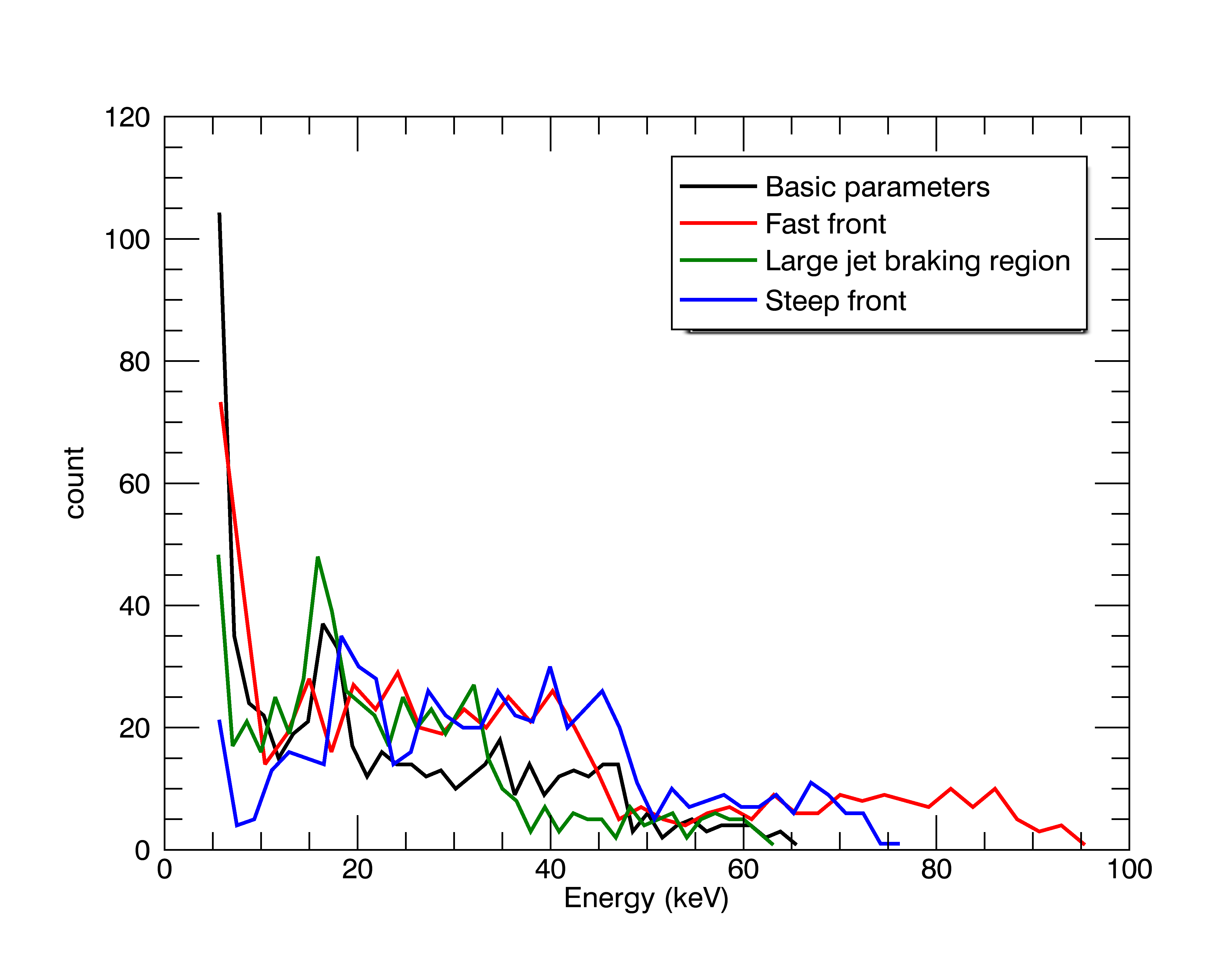}
	\caption{Particle orbit energy frequency for trap parameters given in Table \ref{param}. The maximum kinetic energy of each particle orbit is counted.}
	\label{hist}
\end{figure}

The results of our test runs are shown in Figures \ref{dataset} and \ref{hist}, as well as in Tables \ref{freq} and \ref{max-ke}. Figure~\ref{dataset} illustrates how the results shown in Figure~\ref{explanation}b change when parameters are varied according to the categories in Table \ref{param}. Figure~\ref{hist} shows histograms of the number of particle orbits per given energy. Table \ref{freq} counts the number of times at which each type of orbit is observed in each test run. Table \ref{max-ke} shows the maximum energy obtained by any orbit for each orbit classification in each test run, as well as the average of the maximum energy values.

Figure~\ref{dataset}a shows the average test particle orbit positions from Figure~\ref{explanation}b, where the maximum energy of each orbit defines the colour of each point. The highest energies are achieved by particle orbits of type 2a, followed by type 2b and type 1. Type 2b orbits show higher average maximum energies than type 1, although the most energetic orbit in both categories achieves a kinetic energy of approximately 50 keV. This indicates that type 1 orbits are not as efficient at energizing particles as type 2 orbits are. Type 2a orbits are similar to that shown in Figure~\ref{high-energy} (in that they are trapped between mirror points in regions a and b in Figure~\ref{mirror}). In contrast, type 2b orbits are similar to those started at $52.6\unit{Mm}$ and $49.5~\unit{Mm}$ from Figure~\ref{particle2}, and are trapped further down the loop leg. 

\begin{table}
\begin{tabular}{c l l}
	\hline
					& Parameters	& Initial conditions \\
	\hline
		Figure~\ref{dataset}a 	& Basic		& IC1\\
	\hline
		Figure~\ref{dataset}b	& Basic except $v_\phi = -3, \sigma = 3$	& IC2 \\
	\hline
		Figure~\ref{dataset}c	& Basic except $d = 0.8, w = 0.5$	& IC3 \\
	\hline
		Figure~\ref{dataset}d	& Basic except $\a = 2.0$	& IC1 \\
	\hline
\end{tabular}
\caption{Trap parameters and test particle orbit initial conditions used to examine the effect of changes in trap parameters on particle orbits. The basic parameters are given in Table \ref{parameters}.}
\label{param}
\end{table}

Increasing the initial flow velocity of the braking jet from its value in the basic parameters (see Table \ref{param}) accelerates test particle orbits to higher energies, as can be seen in Figure~\ref{dataset}b, and from the red curve in Figure~\ref{hist}. Table \ref{max-ke} shows increases in the maximum and average kinetic energy obtained by the orbits, with the most energetic particles reaching up to $95.2~\unit{keV}$ in comparison to $65.4~\unit{keV}$ obtained with the basic trap parameters. Figure~\ref{dataset}b shows a new subset of type 2 particles with $|\overline x| > 10\unit{Mm}$ and $\overline y < 20\unit{Mm}$, which we refer to as type 2c (labeled in Figure~\ref{dataset}b). Type 2c orbits correspond to particles trapped very low in the loop legs and also show high energy gains. These orbits remain confined in the loop leg at the end of the simulation (between mirror points b and c in Figure \ref{mirror}, albeit with mirror point b being located very close to the location of mirror point c), after the jet front has dissipated, rather than having access to a large portion of the loop. This is due to the increased compression of the lower loops caused by the faster jet front. A rebounding of the lower loops after the dissipation of the front would make these particle orbits much more similar to those of type 2b, as they would no longer be confined to a small region in one of the loop legs. In addition to a new subcategory of particle orbits, an increase in initial flow velocity results in the particle orbit categories being more spread out in terms of average position. This indicates particle orbit categories are more distinct, with less overlap between categories. There is a general increase in $\overline y$ for particle orbits of type 3 (seen in Figure~\ref{dataset}b). This is due to the orbits and the braking jet being initialized at a higher position so that the jet front would have more time to come to a stop. For orbits which remain trapped throughout the simulation $\overline y$ does not change as much because $\sigma$ is adjusted so that the front stops at a similar height as in the basic parameter case. Table \ref{freq} shows that a faster front is more efficient at trapping particles, particularly in type 2 orbits, in comparison to a slower front.

Increasing the size of the jet braking region results in less distinction between orbits of type 2a and 2b (see Figure~\ref{dataset}c). The maximum energy achieved by type 1 particles is also reduced by a larger jet braking region; the energy gains made by all other orbits remain relatively unchanged from those seen in the case with basic trap parameters (as can be seen in Table \ref{max-ke}).

Finally, increasing the steepness of the front caused by the braking jet results in a stronger magnetic field in the jet braking region. The stronger magnetic field in the braking jet region causes trapping of particle orbits in this region to become much less frequent (see Table \ref{freq}), which results in type 2 orbits becoming more frequent. In particular, all type 2 orbits obtained are of type 2b. We also see a modest rise in maximal energy of type 2 orbits compared to the case with basic trap parameters. The maximum energy achieved by a type 1 orbit is lower, only $38.6~\unit{keV}$, in comparison to $50.9~\unit{keV}$ for the basic parameter case. The average of the maximum energies of type 1 orbits is slightly higher ($22~\unit{keV}$ compared to $18.1~\unit{keV}$ for the basic parameters). Type 3 orbits show small increases in energy (both maximum and average over all type 3 orbits) in comparison with the basic parameter case.  

In each of the cases outlined above the populations of particle orbits generated depend both on the magnetic field structure of the CMT and also on the initial conditions of the test particle orbits. The structure and evolution of the magnetic field is of great importance for the possibility of different populations of particle orbits occuring. For instance, the presence of different subcategories of type 2 orbits depends on the velocity and shape of the front. For slower jet fronts it is not possible to obtain populations of particle orbits of type 2c. Conversely, increasing the steepness of the front causes the type 2a orbits to disappear, leaving type 2b orbits the dominant behaviour in this test run. The initial conditions of the test particles also strongly influences particle orbit behaviour. For an individual particle orbit, small changes in initial conditions (such as its initial position with respect to the front) can result in significantly different behaviour. For example, for a given pitch angle, horizontal displacement of the initial position can change the field line orbited by the particle, together with the location of possible mirror points. This may cause particles to change trapping behaviour. These findings suggest that it may be very difficult to predict the behaviour of a given particle orbit based on initial conditions alone, particularly when trying to differentiate between different subcategories of type 2 orbits.

The effect of different parameter regimes on test particle orbit energies is shown in a histogram of maximum orbit energies for each of the runs discussed previously (see Figure~\ref{hist}). In comparison to the basic parameters, Figure~\ref{hist} shows that increasing the speed of propagation of the jet front causes an increase in the number of high energy orbits. A larger jet braking region causes more trapping and accelerates more particle orbits to modest energies in the range of $10~\unit{keV}$ to $20~\unit{keV}$, however fewer orbits gain energies beyond $35~\unit{keV}$. Finally a steeper front produces a dramatic increase in test particle orbits with energies in the range of 15 to $50~\unit{keV}$. Few orbits are left that have energies less than $15~\unit{keV}$. 

Our results may be explained as follows. A faster jet results in more pileup of magnetic flux at the jet front, producing more energy gain and altering the trapping characteristics to favour type 2 motion more than with a slower jet. A larger braking jet region results in more trapping, however with no increase in magnetic field strength, the orbits do not gain as much energy as in the case of the faster jet or the steeper front, yielding many orbits with small energy gains. An increase in the steepness of the front results in a significantly greater magnetic field strength, which produces much less trapping in the jet braking region. This is because the magnetic field strength is more uniform between the centre of the trap and region ``a'' in Figure~\ref{mirror}. In this scenario test particle orbits are more efficiently trapped in the loop legs, resulting in type 2 orbits.

\begin{table}
\begin{tabular}{c c c c c c}
\hline
	& Basic 	& Fast Front 		& Large jet braking region 	& Steep Front 	\\
\hline
type 1	& 78 	& 52		& 99	& 21	\\
type 2a	& 75	& 66		& 362	& 	\\
type 2b & 260	& 205 		&	& 497	\\
type 2c & 	& 127		&	&	\\
type 3	& 187	& 150		& 139	& 82	\\
\hline
\end{tabular}
\caption{Particle orbit frequencies for different test runs.}
\label{freq}
\end{table}

\begin{table}
\begin{tabular}{*{5}{c}}
\hline
Orbit type	& \multicolumn{2}{c}{Basic} 	& \multicolumn{2}{c}{Fast front} 	\\
		& maximum & average 		& maximum & average 			\\
\hline
1	& 50.9 	& 18.1	& 73.1	& 17.5	\\
2a	& 65.4 	& 45.4	& 95.2	& 74.5	\\
2b 	& 49.4 	& 29.4	& 86.2 	& 34.6	\\
2c 	&  	&	& 93.3 	& 49.1	\\
3	& 17.2 	& 8.9	& 21.7	& 11.1	\\
\hline
Orbit type	& \multicolumn{2}{c}{Large jet braking region} 	& \multicolumn{2}{c}{Steep front} 	\\
		& maximum & average 				& maximum & average 			\\
\hline
1	& 34.5	& 17.2 	& 38.6 	& 22.0	\\
2a	& 63.0	& 31.1 	&  	& 	\\
2b 	&	&	& 76.0	& 39.8	\\
2c 	&	&	&	&	\\
3	& 16.9 & 10.5	& 19.0 	& 12.4	\\
\hline
\end{tabular}
\caption{Maximum and average kinetic energies (in keV) obtained for each type of particle orbit for parameter values and initial conditions shown in Table \ref{param}. Average energies are calculated by averaging the maximum energies achieved by each particle orbit for each orbit type.}
\label{max-ke}
\end{table}

%%%%%%%%%%%%%%%%%%%%%%%%%%%%%%%%%%%%%%%%%%%%%%%%%%%%%%%%%%%%%%%%%%%%%%%%%%%%%%%%%%%%%%%%%%%%%%%%%%%%%%%%%%%%%%%%%%%%%%%%

\section{Conclusions}

We have identified new types of trapping in CMTs with the addition of a braking jet, which (crucially) cannot be obtained using earlier models. The braking jet allows trapping of test particle orbits in the central jet region, in the loop legs and a combination of the two. The regions accessible to orbits are determined by the initial conditions (such as initial position with respect to the front, initial pitch angle and the trap parameters). Particle orbit energy gains also depend on the region which traps them, with orbits which are trapped in the loop legs able to gain more energy than those trapped in either the jet braking region, or those which escape the trap within the simulation time.

We have also shown that variations in trap parameters can greatly impact particle orbits and energy gain. As shown in \cite{artemyev2014}, increasing the speed of the braking jet increases the maximum energy that particles may achieve. It is possible to obtain energies as high as $100~\unit{keV}$ for parameter choices reflective of conditions in the solar corona (for instance, we used an initial jet speed of $3\times 10^6~\unit{m s}^{-1}$, length scale $10^7\unit{m}$ and magnetic field strength scaling with $0.01\unit T$, although values for the magnetic field initially within the braking jet are lower.). Higher initial jet velocities yield larger differences between the different types of particle orbit and allows trapping of particles very low in the loop legs. Larger braking jets show more trapping and hence a larger fraction of test particle orbits get accelerated, but only to moderate energies. We also find that a steeper gradient in the magnetic field as the front propagates yields significantly different behaviour. Trapping of test particle orbits in the jet braking region becomes less frequent due to the higher magnetic field strength throughout the front. As a result more test particle orbits are trapped in the loop legs. The maximum energies obtained this way are higher than in the case without a steep gradient, but not as high as for the faster initial jet. 

Further investigation of particle acceleration in this model is possible by varying more trap parameters. As mentioned in Section \ref{frequencies}, one interesting extension of the investigation presented in this paper would be to model a rebounding of the loops compressed by the plasma jet, leading to a different final position of the trapped particle population.

%%%%%%%%%%%%%%%%%%%%%%%%%%%%%%%%%%%%%%%%%%%%%%%%%%%%%%%%%%%%%%%%%%%%%%%%%%%%%%%%%%%%%%%%%%%%%%%%%%%%%%%%%%%%%%%%%%%%%%%%

\appendix
\section{Description of Transformation}

\begin{figure}[h!]
\centering
	\includegraphics[width = 0.7\textwidth]{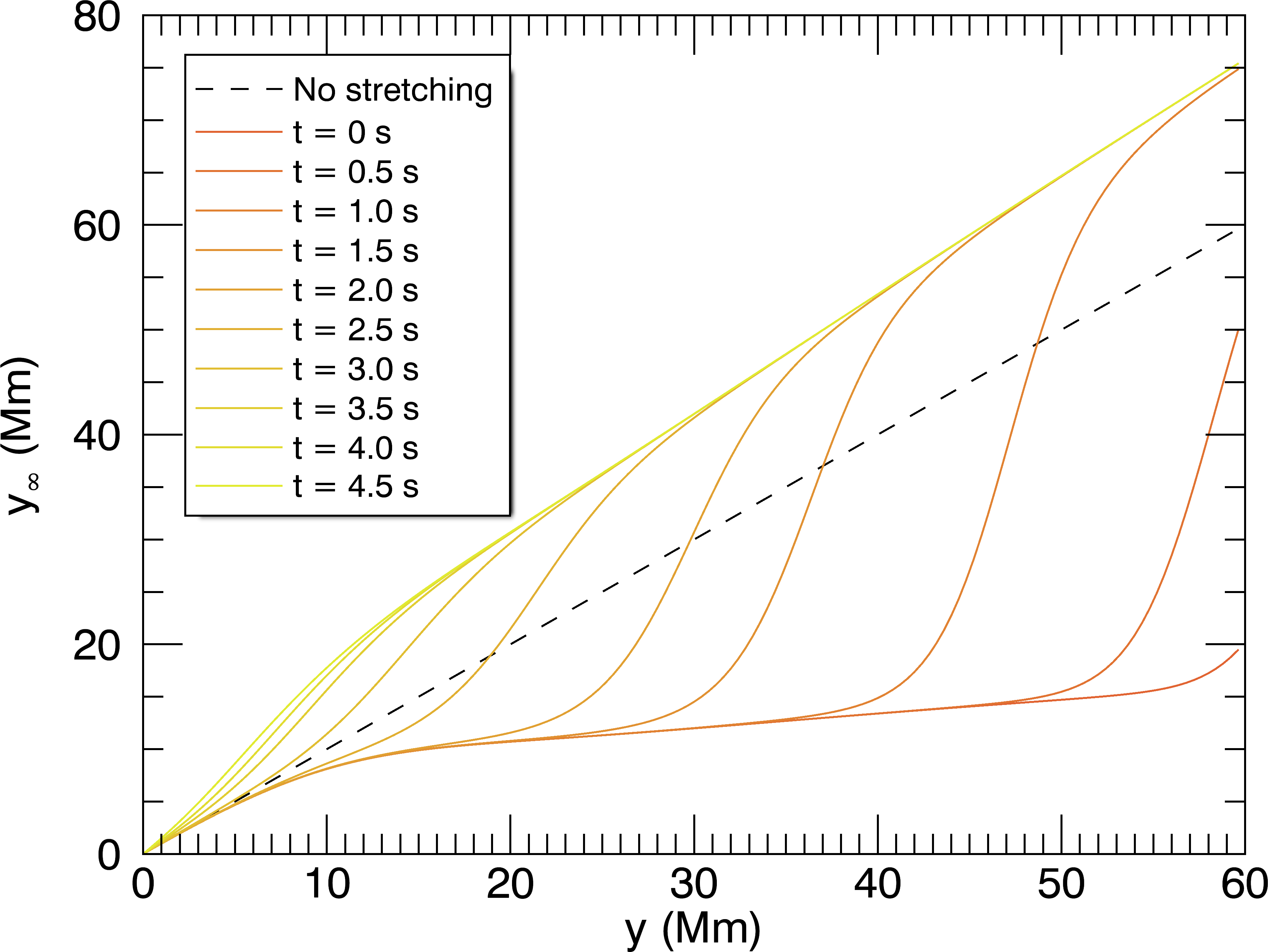}
	\caption{Transformation with $x = 0\unit{Mm}$ (in centre of CMT). Regions for which $y_\infty < y$ correspond to stretching, while regions with $y_\infty > y$ correspond to compression. For a fixed time, the front is located in the region where $y_\infty$ increases rapidly. The dashed line is a visual aid, set at $y_\infty = y$.}
	\label{t-idea}
\end{figure}

 Here we discuss how the transformation defining the analytical fields is obtained. For a detailed discussion of the theory behind the transformation method see \cite{giuliani-et-al2005}. We start with an initial stretched configuration similar to that in \citet{giuliani-et-al2005} to which a front with a sharp increase in magnetic field strength is added. Behind the front the field lines are closer to their equilibrium configuration, similar to what is suggested in \cite{artemyev2014}. The $y$-component of the transformation used in this paper, restricted to $x = 0\unit{Mm}$ is shown in Figure~\ref{t-idea} for multiple values of $t$.

Vertical stretching corresponds to $y_\infty < y$, whereas for compression $y_\infty > y$. Figure~\ref{t-idea} shows that when $t = 0\unit{s}$ and $y < 60~\unit{Mm}$, $y_\infty < y$, which corresponds to a vertical stretching. As the front passes a given $y$ the value of $y_\infty$ increases rapidly, indicating compression caused by the front. The transformation shown in Figure~\ref{t-idea} is given by

\begin{equation}\label{t-propose}
	y_\infty = S + y\frac{1 + \tanh \phi}{2},
\end{equation}
where the first term, $S$, defines the shape of the field before the front has passed. The second term describes the shape of the front, where $\phi = 0$ is the location of the front. The function $\phi$ depends on position and time. It satisfies $\phi < 0$ before the front passes a particular location and $\phi > 0$ after the front has passed. 

One choice for the first term of Equation~(\ref{t-propose}) that produces the desired vertical stretching is 
\begin{eqnarray}
	\hspace{-0.4cm} S &=& s\log \LL 1 + \frac{y}{s} \RR  \LL 1 - \frac{1 - \tanh(y - y')}{2}\cdot \frac{\tanh(x + x') - \tanh(x - x')}{2} \RR \\
	        && + y \LL \frac{1 - \tanh(y - y')}{2}\cdot \frac{\tanh(x + x') - \tanh(x - x')}{2} \RR. \nonumber
\end{eqnarray}
This choice imposes a stretching of the form $s\log \LL 1 + y/s \RR$ for points outside the box given by $0 \leq y \leq y'$ and $-x' \leq x \leq x'$, and leaves the interior untransformed (\textit{i.e.} in its final configuration). We currently do not confine the box in the $x$ direction by setting $x' = 100$ (outside of the region we investigate in this paper). 

To choose a functional form for $\phi$ we consider the front position determined by
\begin{equation}\label{phi-try}
	\phi = y - v_\phi \sigma \tanh\LL t/\sigma \RR - y_0.
\end{equation}
The location of the front is given by $\phi = 0$. For small values of $t$ the location of the front is $y = y_0 + v_\phi \sigma \tanh \LL t/\sigma \RR \simeq y_0 + v_\phi t$, which means that the front propagates with a constant speed. For larger values of $t$, $\tanh\LL t/\sigma \RR$ approaches a constant, meaning the front slows down and stops. Calculating 
\begin{equation}
\lim_{t \to \infty} \LL y_0 + v_\phi \sigma \tanh \LL t/\sigma \RR\RR = y_0 + v_\phi \sigma
\end{equation}
shows that the parameter $\sigma$ controls how deeply the front penetrates into the equilibrium loops at the bottom of the trap because it determines the final location of the front.  

It is possible to modify the steepness of the front (which in turn affects the strength of the electric and magnetic fields at the front) by increasing the gradient of $\phi$. We achieve this by multiplying the expression given in Equation~(\ref{phi-try}) by the factor $\a\LL \LL y + 1 \RR \LL s_{o} x^{2} + 1 \RR  \RR ^\b$. The factor $(y+1)^\b$ causes the transformation to be steeper for larger values of $y$. Multiplication of $\phi$ by the factor $(y+1)^\b$ changes the shape of the loops, so the factor $\LL s_o x^2 + 1 \RR^\b$ is added to correct for this effect. The functions $J$ and $T$ (given in Equation~(\ref{modified-jet-trap})) are added to further modify the shape of the CMT:
\begin{equation}\label{modified-jet-trap}
	J = d \ex^{-x^2 y /w}, \qquad T = k \tan \LL \frac \pi 2 \frac{x^2}{w_2} \RR \tanh y.
\end{equation}
The function $J$ produces an indentation in the jet braking region. The indentation takes the form of an exponential with its depth and width determined by the parameters $d$ and $w$. The function $T$ modifies the shape of the trap for large values of $|x|$ to maintain the shape of the loops. 

\citet{artemyev2014} suggests that as the braking jet propagates the front may become steeper. In the transformation described above the front becomes shallower as the jet propagates towards the solar surface. The reason for this is that as the braking jet approaches the lower loops it is travelling slow enough that the magnetic flux that piled up previously spreads out. The spreading out of the magnetic flux causes the front to become less steep and eventually disappear. Nevertheless, closer to the reconnection region the outflow is faster and the magnetic field not as strong so we expect to see a steepening front. To incorporate this steepening into our model $\phi$ and $T$ are multiplied by the factor $ 1 + \frac{1}{2}\LS \chi \sin\LL \pi y/y_0 \RR - 1 \RS\LS 1 - \tanh(\zeta (t - t')) \RS $. This corresponds to multiplying $\phi$ and $T$ by $\chi \sin\LL \pi y/y_0 \RR$ when $t \ll t'$, producing a steepening of the transformation near $y = y_0$. The result is the transformation presented in Equations~(\ref{yinf}) -- (\ref{trap}). 

%%%%%%%%%%%%%%%%%%%%%%%%%%%%%%%%%%%%%%%%%%%%%%%%%%%%%%%%%%%%%%%%%%%%%%%%%%
% Acknowledgements

\begin{acks}
AB would like to thank the University of St Andrews for financial support from the 7th Century Scholarship and the Scottish Government for support from the Saltire Scholarship. TN and JT gratefully acknowledge the support of the UK Science and Technology Funding Council (Consolidated Grant ST/K000950/1).

\end{acks}

%% %%%%%%%%%%%%%%%%%%%%%%%%%%%%%%%%%%%%%%%%%%%%%%%%%%%%%%%%%%%
% Bibliography

%Using BibTeX

\bibliographystyle{spr-mp-sola}
\bibliography{references}

%Without BibTeX
%\begin{thebibliography}{}
%\bibitem[\protect\citeauthoryear{Author}{Year}]{key}
%  <bibliographical entry>
%
%\bibitem[\protect\citeauthoryear{}{}]{}
%
%
%\end{thebibliography}

\end{article}
\end{document}